\documentstyle[12pt,epsf]{article}

\newcommand{\be}{\begin{equation}}
\newcommand{\ee}{\end{equation}}
\newcommand{\ba}{\begin{eqnarray}}
\newcommand{\ea}{\end{eqnarray}}

\newcommand{\GT}{{O(4,4;{\bf Z})}}
\newcommand{\GU}{{O(5,5;{\bf Z})}}
\newcommand{\cM}{{{\cal M}_{BPS}}}
\newcommand{\bM}{{\bar M}}
\newcommand{\cF}{{\cal F}}

\newcommand{\bOmega}{{\bar \Omega}}

\newcommand{\non}{\nonumber\\}
\newcommand{\del}{\partial}
\newcommand{\dis}{\displaystyle}

\newcommand{\nom}{\nonumber}

\newcommand{\dsp}{\displaystyle}
\newcommand{\eqn}[1]{(\ref{#1})}
\newcommand{\br}{\mbox{$\bf R$}}

\newcommand{\bz}{\mbox{$\bf Z$}}

\newcommand{\df}{\stackrel{\rm def}{=}}

\newcommand{\de}{\delta}

\newcommand{\al}{\alpha}
\newcommand{\la}{\lambda}

\newcommand{\lb}{\lbrack}
\newcommand{\rb}{\rbrack}

\newcommand{\ca}[1]{{\cal #1}}
\newcommand{\msc}[1]{\mbox{\scriptsize #1}}
\newcommand{\ket}[1]{{|#1 \rangle}}
\newcommand{\bra}[1]{{\langle #1|}}
\newcommand{\tri}{\mbox{\large $\triangle$}}
\newcommand{\id}{\mbox{\bf{1}}}

\newcommand{\dil}{e^{-\Phi/2}}
\newcommand{\cm}{{\cal M}}
\newcommand{\cleqn}{\setcounter{equation}{0}}

\begin{document}

\begin{titlepage}
\nopagebreak

\vfill
\begin{center}
{\LARGE D-brane Analyses for BPS Mass Spectra}

~

{\LARGE and U-duality}

\vskip 12mm

{\large Y.~Sugawara${}^{\sharp}$ and
K.~Sugiyama${}^{\S}$
}
\vskip 10mm
${}^{\sharp}$
{\sl  Department of Physics, Faculty of Science, University of Tokyo}\\
{\sl  Hongo, Bunkyo-ku, Tokyo 113-0033, Japan}\\
\vskip 2mm
${}^{\S}$
{\sl Department of Fundamental Sciences}\\
{\sl Faculty of Integrated Human Studies, Kyoto University}\\
{\sl Yoshida-Nihon-Matsu cho, Sakyo-ku, Kyoto 606-8501, Japan}\\
\end{center}
\vfill

\begin{abstract}
We give a confirmation of U-duality of type II
superstring by discussing mass spectrum of the BPS
states.  
We first  evaluate the mass spectrum of BPS solitons with 
{\em one kind of\/} R-R charges. 
Our analysis is based on the 1-loop effective action
of D-brane, which is known as ``Dirac-Born-Infeld (DBI) action'', 
and the fact that BPS states correspond to the SUSY cycles with
minimal volumes. We show the mass formula 
derived in this manner  is completely fitted with that given by U-duality. 
  
We  further discuss the cases of BPS solitons possessing {\em several kinds 
of\/} R-R charges. These are cases of ``intersecting
D-branes'',
which cannot be described by simple DBI actions. 
We claim that, in these cases, higher loop corrections should be incorporated  
as binding energies between the branes. It is remarkable that the summation
of the contributions from all loops reproduces the correct mass formula 
predicted by U-duality.
\end{abstract}
PACS codes; 11.10.Kk, 11.25.-w, 11.25.Sq, 11.30.Pb\\ 
Keywords; D-brane, BPS state, U-duality, mass formula, M-theory\\

\end{titlepage}

\section{Introduction}

D-brane analyses have  been found to be very successful
methods to describe the solitonic states of the
string theory \cite{DLP}-\cite{BVS2}.
They  give essential insights into the non-perturbative
aspects of the string theory, such as the recent
success of the microscopic description of the
black hole properties \cite{DGHR}.
By the use of string theory, they give a hope that
those non-perturbative excitation might be quantized.

In this paper, 
we focus on the type II superstrings compactified on torus.
This theory is conjectured to have the U-duality
symmetry \cite{Se2,HT,SV} which originates from
the symmetry of the supergravity theory \cite{CJ}.
Our aim is to present a non-trivial check of  U-duality.
One of the remarkable successes to this aim is the calculations of
degeneracy of BPS states given in the excellent works \cite{V2,BVS2}.
However, these calculations are essentially independent 
of the background moduli. Hence it is still meaningful 
to study the quantities which strongly depend on the moduli    
for the confirmation of U-duality. One of the typical objects
with this property is the BPS mass spectrum.

Motivated with this fact,
we intend to examine the mass formula for
BPS solitons with Ramond-Ramond (R-R) charges 
by means of D-brane techniques. We shall compare it 
with that predicted by U-duality.

For the simple cases in which only one kind of R-R charges are excited,
we shall analyse the mass spectra 
by using the Dirac-Born-Infeld (DBI) action for the D-brane  
\cite{L,CK,Town,Sch}.
This is an extension of the study in our previous
publication \cite{IMSS}.

It is further stimulating  to analyse the more complicated 
situations with several kinds of R-R charges excited, that is, 
the R-R solitons described by 
the ``intersecting D-branes'' \cite{D,Se2,BVS2,PT,BL}. 
We will stress that the analysis based on DBI action, 
which includes only the one-loop contribution, 
is {\em not\/} sufficient to discuss the D-brane bound states. 
Under general backgrounds, the configurations of intersecting D-branes
break SUSY completely. In those cases, the (stringy) higher loop corrections 
do not vanish. 
We will show that the higher loop corrections can be regarded as 
the binding energies among the intersecting branes and analysis based
on these loop corrections give 
the consistent results with U-duality.

~

Let us give the plan of this paper.
In section 2,
we give the BPS mass formula which is invariant under
U-duality by extending the well-known
BPS spectrum of the fundamental string sector \cite{GPR}.

In section 3  we consider the simple BPS solitons 
possessing only one kind of R-R charges.  
The BPS states of this type are realized as 
the supersymmetric cycles \cite{BBS,HL},  
equivalently, geometrical configurations with
minimal volumes. We evaluate the D-brane mass
directly from the DBI action by
constructing such a configuration.
It is an important check 
of U-duality to study the objects depending strongly on 
the background moduli, 
because the duality transformations map the moduli in a string theory
onto those in a dual string theory non-trivially. 
We will show that the DBI action
produces the mass spectra with correct moduli dependences
expected by U-duality. In the type IIA case, we also explain the BPS
mass formulae from the M-theory 
viewpoint. It is based on SUSY
algebra \cite{susyDVV} 
and the results confirm the validity of our analysis in the
DBI action.

We will also argue on the situations in which 
the gauge fields on the world volumes of branes have topological charges. 
It is believed that these charges are induced by
sub-branes within other D-branes (``{\em branes within branes}'' \cite{D}).
We confirm that our evaluation of BPS masses based on the DBI action 
produces the results which are consistent with this interpretation.
That is, we show that the calculation under the non-trivial background
gauge fields yields the correct masses of the bound states expected 
by U-duality. 

Although this result is very satisfactory,
we should still keep in mind the fact that 
this story is not complete to understand 
the physics of D-brane bound states. 
There still exist many bound states which cannot be
reduced to the cases of ``branes within branes''.
It is sufficient to take only an example to illustrate such generic
situations; the 1-branes wrapping 
around 9th-axis and the 3-branes wrapping around
678th-axes (we assume the 6789th-directions correspond to the compactified
4-torus). This situation and its mass formula cannot be studied
by the trick  of gauge fields.

To defeat this difficulty 
we investigate, in section 4,  the problem of bound states 
not only in the ``branes within branes'' cases but also in 
the situations that some branes ``literally'' intersect with others.
We emphasize the necessity of  taking 
higher loop corrections into account in the latter cases. 
We discuss relations between the SUSY breaking 
{\em under generic moduli\/} and the non-vanishing 
higher (string) loop amplitudes associated with annulus diagrams. 
We will evaluate the binding energies of 
intersecting branes as contributions from higher loops,
and show a remarkable fact: The summation of all loop corrections 
reproduces the correct mass formula of bound states predicted by 
U-duality!
In the last, we will try to explain the possibility to describe 
some ({\em not all\/}) bound states
by only DBI actions from the point of view of
``{\em geometrization of quantum correction}''.

The last section is devoted to conclusion and 
a few  comments on the open problems.

~

~

\section{BPS Mass Formulae with U-duality Invariance}\label{U-mass}
\cleqn

In this section, we shall analyse the BPS mass formulae in the IIB
superstring compactified on $T^4$. The NS-NS sector mass formula can
be written in a manifestly T-duality invariant form. Then we can
derive the BPS mass formula in the R-R sector by using a (special)
U-duality transformation.


Let us consider type IIB superstring compactified on $T^4$.
The massless states in the type IIB string are
metric field $G_{MN}$, second order antisymmetric tensor
$B_{MN}$, and dilaton $\phi$
from NS-NS sector. A scalar field (axion) $C^{(0)}$, 
an antisymmetric tensor $C^{(2)}_{MN}$, and a self-dual 
fourth order antisymmetric tensor $C^{(4)}_{MNPQ}$ appear in the
R-R sector.

After the compactification,
the scalar fields which describe the moduli of the theory
are given by
$G_{ij}$,
 $B_{ij}$ and dilaton $\phi$
 from NS-NS sector.
We use indices $i,j$ to run $6,7,8,9$
to describe the internal space coordinates and use $\mu, \nu$
for uncompactified dimensions.
$G_{ij}$ and $B_{ij}$ give 
a structure of Grassmannian,
$M\in O(4,4)/(O(4)\times O(4))$ to the moduli by a combination,
\ba\label{mass_fund}
M&=&
(\Omega_{B}\Omega_{e})(\Omega_{B}\Omega_{e})^t\,\,\,,\,\,\,
\Omega_e=\left(
\begin{array}{cc} e & 0\\ 0 & (e^t)^{-1}
\end{array}
\right)
\,\,\,,\,\,\,\Omega_B=\exp\left(
\begin{array}{cc} 0 & B \\ 0 & 0
\end{array}\right).
\label{OmegaB}
\ea
The $e$ is the vierbein of the string (sigma model) metric $G$, 
namely, $G=e\cdot e^t$.
$\Omega_{*}$ belongs to $O(4,4)$ and satisfies a relation
\be
\Omega_{*} J \Omega_*^t = J
\,\,,\,\,
J=\left(
\begin{array}{cc}
0 & I_4\\
I_4 & 0
\end{array}
\right).\nonumber
\ee
Here $I_4$ is a $4\times 4$ unit matrix.
The left action of $\varpi \in \GT$
($\Omega_*'=\varpi\Omega_{*}$)
represents the
T-duality of the system.
On the other hand, the Ramond moduli,
$C^{(0)}$, $C^{(2)}_{ij}$ and $C^{(4)}_{6789}$ 
are combined to give an 8 component
field
$\psi^{(\alpha')}$ ($\alpha' = 1,2,\cdots , 8$) 
in the cospinor representation of $O(4,4)$,
which we write as $(\psi')^{(\alpha')}
=R_c(\varpi )_{\alpha' \beta'}\psi^{(\beta')}$.
(Similarly, in the type IIA case, 
$C^{(1)}_i$ and $C^{(3)}_{ijk}$ are combined into a spinor multiplet
of $O(4,4)$ ).

The vector fields in 6 dimensions are essentially composed of  
$G_{i\mu}$ and $B_{i\mu}$. The former is 
associated with the Kaluza-Klein momentum
and the latter is coupled with the winding number around $i$-th direction.
The NS vector fields are combined into a
single 8 component gauge field $A_\mu^{(a)}$
($a=1,2,\cdots ,8$) in
a vector representation under $O(4,4)$, ${A'}_\mu^{(a)}=
\varpi_{ab}A^{(b)}_{\mu}$. 
Vector fields from the Ramond sector
$C^{(2)}_{i\mu}$, $C^{(4)}_{ijk\mu}$
count the 1,3-brane charges. They
are combined into an 8 component vector
field $K_\mu^{(\alpha)}$ which transforms as
a spinor under $O(4,4)$, $K_\mu^{(\alpha)}\rightarrow
(R_s(\varpi))_{\alpha\beta}K_\mu^{(\beta)}$.
These spinor representation
matrices, $R_{s}(\varpi)$ , satisfy
$R_{s}(\varpi)JR^t_{s}(\varpi)=J$.
(More precisely, there appear  some mixing 
of $G_{i\mu}$ and $B_{i\mu}$ in the definition of $A_\mu^{(a)}$
and that  of $C^{(i)}$'s in the definition of $K_\mu^{(\alpha)}$
under general backgrounds.)

We write the integral
charges $n^{(a)}, \, m^{(\alpha )}$ associated with
$A_\mu^{(a)}$ and $K_\mu^{(\alpha )}$.
These charges transform as
vector ($n^{(a)}$) and spinor ($m^{(\alpha )}$) of $\GT$.
For each set of integers, we can define  a stable state called 
the BPS state.

In the fundamental string spectrum, we have
vanishing R-R charges,  $m^{(\alpha )}=0$.
The famous mass formula of the (anti)BPS state
in this case is given as \cite{GPR},
\ba
\cM^2
&=&{\bf n}^t (M \pm J) {\bf n}
={\bf n}^t  (\Omega_B\Omega_e)\Pi^\pm (\Omega_B\Omega_e)^t {\bf n}\,\,\,,
\label{eqn:tmass}\\
\Pi^\pm &=&\left(
\begin{array}{cc}
I_4 & \pm I_4\\
\pm I_4 & I_4\\
\end{array}
\right).\nonumber
\ea
Here we write the $O(4,4;{\bf Z})$ vector
${\bf n}=(n^{(a)})=\left(\begin{array}{c}
w^i \\ n_i \end{array}\right)$, 
($n_i;$ KK momentum, $w^i;$ winding number). 
The $n^{(a)}$ couples with the gauge field $A^{(a)}_{\mu}$
as ${\bf n}^t A_{\mu}$.
The $\Pi^\pm$ are  appropriate operators
that project to
(anti)BPS states.
Needless to say, this mass formula is invariant under the right action of
$\sigma\in O(4)\times O(4)$
since $\Omega_B\Omega_e$ is transformed into
$(\Omega_B\Omega_e)\sigma$ and
$ \sigma \Pi^\pm \sigma^t =\Pi^\pm$.
The invariance under T-transformation ($O(4,4;{\bf Z})$ left-action) 
is ensured  by the transformation law;
\be
\left\{ \begin{array}{lll}
    \Omega_B \Omega_e & \longrightarrow &  \varpi(\Omega_B \Omega_e)\,, \\
    {\bf n} & \longrightarrow & (\varpi^t)^{-1} {\bf n} 
                     (\equiv J\varpi J {\bf n})\,,
\end{array}\right. ~~~ ({}^\forall \varpi \in O(4,4 ; \bz))\,.
\ee

We shall make here one important remark: The mass formula \eqn{eqn:tmass}
is written with respect to the sigma model metric $G_{\mu\nu}$.
Later we will observe that $G_{\mu\nu}$ is not invariant under 
general U-duality transformations (even though it is invariant under 
all the T-transformations $O(4,4 ; \bz)$).
Therefore we should write down the mass formula 
associated {\em not\/} to the sigma model metric $G_{\mu\nu}$  
{\em but\/} to the 6-dimensional Einstein frame metric 
$g^{(6)}_{\mu\nu} \equiv e^{-\Phi}G_{\mu\nu}$
which is U-invariant.
Here $\Phi$ denotes the 6-dimensional dilaton 
that is invariant under T-transformations;
\ba
\Phi = \phi - \frac{1}{4} \log \det (G_{ij}).\label{6dim dilaton}
\ea
It is easy to observe that the mass formula defined with respect to
$g^{(6)}_{\mu\nu}$ should be related with that for $G_{\mu\nu}$  
\be
{\cal M}^{\msc{(Einstein)}}=e^{\Phi /2} {\cal M}^{\msc{(sigma model)}}.
\ee
Namely, we should rewrite the mass formula for the NS-NS charges
\eqn{eqn:tmass} as 
\ba
\cM^2 &=& 
e^\Phi {\bf n}^t  (\Omega_B \Omega_e) 
  \Pi^\pm (\Omega_B \Omega_e)^t {\bf n}\non
&=&e^\Phi {\bf n}^t ( M \pm J) {\bf n}\,\,\,.
\label{Umass1}
\ea


Beside $\GT$, the type IIB superstring theory is conjectured
to be invariant under the strong-weak $SL(2;\bz)$ duality
(S-duality).
This symmetry is combined with T-duality symmetry to
give the U-duality symmetry group $\GU$ \cite{Se2,HT,SV,W2}.

In order to clarify the structure of U-duality $\GU$,
we must suitably arrange the moduli fields from 
both the NS-NS sectors $G_{ij}$, $B_{ij}$, $\Phi$ and the R-R sectors
$\psi_{\alpha}$.
The $\Omega_e,\,  \Omega_B $ should be  embedded in $O(5,5)$-matrix 
as follows (this is one of the conventions);
\be\label{embed}
\bOmega_{e,B}=\left( \begin{array}{c|c} I_2 & 0 \\ \hline 0 &  
R_c(\Omega_{e,B})
\end{array}\right).
\ee
Here the $R_s(\Omega)$ and $R_c(\Omega)$
are spinor and cospinor 
representations of $\Omega\in O(4,4)$. 


The dilaton and RR moduli are also incorporated as
\ba
\bOmega_{e,B,\Phi,\psi}&=&\bOmega_\psi\cdot\bOmega_\Phi
\cdot \bOmega_{B} \cdot \bOmega_{e} \,\,\,,\non
\bM &=& \bOmega_{e,B,\Phi,\psi}\cdot\bOmega_{e,B,\Phi,\psi}^t \,\,\,,
\ea
with
\ba
&&\bOmega_{\Phi}=
\left( \begin{array}{cc|c}
e^\Phi & 0 & 0\\ 0 & e^{-\Phi} & 0\\ \hline 0 & 0 & I_8
\end{array}\right)\,\,\,,\\
&&
\bOmega_{\psi}=\exp\left\{\sum^4_{\alpha' =1}  \psi_{\alpha'} 
(E(2,6+\alpha')-E(\alpha' +2,1))
+\sum^8_{\alpha' =5}  \psi_{\alpha'} 
(E(2,\alpha' -2)-E(\alpha' +2,1))
\right\},
\nonumber
\ea
where $E(i,j)$ is a $10\times 10$ matrix with
only one non-zero entry at $(i,j)$.
Obviously $\bM$ is a symmetric matrix in $O(5,5)$,
and hence parametrizes the ``extended Teichm\"{u}ller space''
$ O(5,5)/(O(5)\times O(5))$.

Now, let us study R-R sector mass formula.
In the following discussion we turn off
all the R-R moduli $\psi_{\alpha}=0$.
We first pick up a special U-transformation
$S^{(6)}:=S^{(10)}T^{6789}S^{(10)}$ (``S-transformation in the sense
of 6-dimension'') such that $(S^{(6)})^2 =\id$.
Here $S^{(10)}$ stands for  the 10-dimensional S-duality transformation
(which corresponds to the transformation 
$\dsp \tau ~\rightarrow ~-\frac{1}{\tau}$ of $SL(2;\bz)$).
Also the $T^{6789}$ is a T-duality transformation along the 6789th
directions.
Then the $S^{(6)}$ acts on the $G$, $B$ and
$\Phi$
\be
S^{(6)}:
\left\{
\begin{array}{lcl}
G_{ij} & \rightarrow & \sqrt{|G|}(G^{-1})^{ij} ,\\
B_{ij} & \rightarrow & \sqrt{|G|}(\ast B)^{ij}
  \equiv {\dsp \frac{1}{2}}\epsilon^{ijkl}B_{kl} ,\\
e^{\Phi} & \rightarrow & e^{-\Phi} ,
\end{array}
\right.
\,\,\,\label{S6-0}
\ee
and completely 
exchanges the NS-NS charges $n^{(a)}$ and the R-R charges $m^{(\alpha)}$.
The 6 dimensional space-time Einstein metric $g^{(6)}_{\mu\nu}$ is
invariant under the $S^{(6)}$, $S^{(10)}$ but the $G_{\mu\nu}$ is not.

It is remarkable that 
\be
S^{(6)}(M) = \left(\begin{array}{cc}
\sqrt{|G|}(G^{-1}-\ast B\cdot G\cdot \ast B) & (\ast B\cdot G)\\
-(G\cdot \ast B) & {\dsp \frac{1}{\sqrt{|G|}}}G  
\end{array}
\right) \equiv R_s(M) \,\,\, , 
\label{S6M}
\ee
where $R_s(*)$ denotes the spinor representation matrix as is already 
introduced. Hence we can rewrite the transformation law \eqn{S6-0} as
\be
S^{(6)}:
\left\{
\begin{array}{ccc}
e^{\Phi}M & \rightarrow & e^{-\Phi}R_s(M) ,\\
e^{-\Phi}R_s(M) & \rightarrow & e^{\Phi}M .
\end{array}
\right.
\,\,\,\label{S6}
\ee
In this way, we can easily find out that 
the mass formula of the R-R sector must take the following form
if we claim the invariance of total mass spectrum under the
$S^{(6)}$-transformation;
\ba
\cM^2 &=& 
e^{-\Phi} {\bf m}^t  R_s(\Omega_{e,B}) \Pi^\pm R_s(\Omega_{e,B}^t)  
                         {\bf m} \non
&=&
e^{-\Phi} {\bf m}^t ( R_s(M) \pm J ) {\bf m} \,\,\,.
\label{Umass2}
\ea
We will compare this formula with the results of D-brane analyses
in the later sections.

~

~
\section{BPS mass formula from Dirac-Born-Infeld action}
\cleqn

In the previous section, we conjectured the BPS mass formula
by using U-duality.  The outcome becomes algebraic
and uses the representation theory of $\GU$.
In the following, on the other hand, we evaluate
the BPS mass  geometrically by minimizing the
D-brane  world-volume, or more precisely  by minimizing  the
Dirac-Born-Infeld type integral.  This  condition should be   
equivalent to
the requirement to keep half of the supersymmetries,
namely the BPS condition.
Such a supersymmetric configuration is called a ``supersymmetric  
$p$-cycle''
\cite{BBS}, and at least
 in the case that Kalb-Ramond moduli
$B_{ij}$ is equal to zero, this statement is proved in  
\cite{BBS}.\footnote{
       For the embedding of two- (resp. three-) cycle into
        the Calabi-Yau two- (resp. three-)fold, this condition
        becomes holomorphic embedding (resp. the special Lagrangian  
submanifold
         \cite{HL}).}
In the general case of $B_{ij} \neq 0$,  the equivalence of the  
conditions
of  minimal volume and supersymmetry is not  so clear. 
But, from the physical point of view 
this assumption is very plausible, since the BPS  
states must respectively  have the minimal energies in the charged  
sectors. It is interesting to observe that
the mass formula obtained geometrically coincides
with algebraic one.

In the subsection \ref{2}, we also  explain the full mass formula in the
IIA theory including both NS-NS and R-R sectors from the M-theory
viewpoint. The analysis is based on the SUSY algebra in
6-dimension \cite{susyDVV}. We will observe that our analysis based on 
the DBI action is consistent with the M-theory approach.

~

\subsection{Type IIB on $T^4$}\label{2b}

First of all, we consider the type IIB case compactified on $T^4$.
For this case the situations are  somewhat  simpler  than those   
for type IIA.

Our  starting point is  the (one loop) effective action of Dirichlet  
$p$-brane
\cite{L,D,Sch} ($p$ is an odd number for type IIB string);
\ba
&& S_{DBI}:=\int   d^{p+1}\sigma\,e^{-\phi}
\sqrt{\det\left(G_{\alpha\beta}+{\cal F}_{\alpha\beta}  
\right)}\,\,\,,  \label{DBI} \\
&&S_{WZ}:=i\, \int   e^{i{\cal F}}\wedge C\,\,\,,  \label{WZ}   
\\
&&\,\,\,\,\,\dsp {\cal F}:=B+\frac{1}{2\pi}F\,\,\,,
\,\,\,C:=C^{(4)}+C^{(2)}+C^{(0)}\,\,\,.\nom
\ea
Here the $G_{\alpha\beta}$, $B_{\alpha\beta}$ are the induced metric
and anti-symmetric field on the world volume. The $C^{(l)}$'s
$(l=0,2,4)$ are the Ramond-Ramond fields for the type IIB string and
$F_{\alpha\beta}$ is the field strength of $U(1)$ gauge field $A$.
$S_{DBI}$ is usually called the ``Dirac-Born-Infeld action''  and
$S_{WZ}$ is often called ``Wess-Zumino term''. (Several authors also  
call it ``Chern-Simons term''.)

We evaluate  the BPS mass for the 
$6$-dimensional  
space-time theory
along the same line of argument developed in the \cite{IMSS}.
Consider  the  $p$-brane with  the  specific configuration ${\bf  
R}\times \Sigma$
($\br$ is the time axis and $\Sigma $ is a $p$-dimensional subspace  
of
the internal $T^4$).
In this setup, the effective mass of  particle  
associated with this $p$-brane is directly evaluated by DBI action; 
\ba
S_{DBI}&=&{\cal M}_p\, \int_{\bf R} \sqrt{g^{(6)}_{00}}d\tau  \nonumber \\
{\cal M}_p &:=& \dil  |G|^{-1/4}
\int_{\Sigma}d^p\sigma\,
\sqrt{\det\left(G_{\alpha\beta}+{\cal F}_{\alpha\beta}\right)}\,\,\,.
\label{mp}
\ea
The $6$-dimensional space-time dilaton $\Phi$
is related with the $10$-dimensional dilaton $\phi$ 
and the torus metric $G_{ij}$ in (\ref{6dim dilaton}), 
in other words
\be
e^{\Phi -\phi} = |G|^{-1/4} ,
\label{dilaton factor}
\ee
where $|G| : = \mbox{det} \, G_{ij}$.
As we already stressed in the previous section, the mass should be
evaluated with respect to the 6-dimensional Einstein frame
$g_{\mu\nu}^{(6)}= e^{-\Phi}G_{\mu\nu}$. We used 
the relation; $\sqrt{g^{(6)}_{00}} \dil |G|^{-1/4}\equiv 
\sqrt{G_{00}}e^{-\phi}$ in Eqs.\eqn{mp}.

The topological nature of $\Sigma$ (more
precisely speaking, the  
homology class of $\Sigma$,
the homotopy class of $X$ and the topological charge of the bundle  
where $A$ is defined)
determines the Ramond-Ramond charge of this particle.
We can find this fact by observing the Wess-Zumino term $S_{WZ}$  
\eqn{WZ} and will see it explicitly in the following examples.
As is already commented,
the BPS state  should correspond to a pair of configurations of the
map $X\, : \,\Sigma \, \rightarrow \,  T^4$ and
the $U(1)$-gauge field $A$ that minimizes the integral \eqn{mp}.
Hence our problem is reduced to search the minimal configuration
of $(X,A)$ under a fixed topological nature of
world volume.

If we turn off the R-R fields, the compactified type IIB moduli space
is described by $G_{ij}$ and $B_{ij}$ of $T^4$.
In the following argument we assume these $G_{ij}$, $B_{ij}$
are all some constants on $T^4$, which is of course possible
since $T^4$ is a flat space.
The induced metric
$G_{\alpha\beta}$ and anti-symmetric tensor $B_{\alpha\beta}$ on the
$p$-brane are defined as
\ba
G_{\alpha\beta}:=G_{ij}\del_{\alpha}X^i\del_{\beta}X^j\,\,\,,\,\,\,
B_{\alpha\beta}:=B_{ij}\del_{\alpha}X^i\del_{\beta}X^j\,\,\,,\nom
\ea
with the  coordinates $\{\sigma^{\alpha}\}$
$(\alpha=1,2,\cdots ,p)$ on the world volume.

~

\noindent
\underline{\em{1-brane (D-string)}}

Let us start by taking  the $1$-brane $({\bf R}\times \Sigma^{(1)})$  
case.
We easily obtain
\ba
{\cal M}_1&=& \dil   |G|^{-1/4} \int_{\Sigma^{(1)}}d\sigma
\sqrt{  G_{\sigma \sigma}}  \nom\\
&\geq &  \dil  |G|^{-1/4}\,  \sqrt{q^iG_{ij}q^j  }    \,\,\,,    
\label{m1}
\ea
where  $\dsp q^i:=\int_{\Sigma^{(1)}}d\sigma\del_{\sigma}X^i $
denotes the winding number of D-string.
Obviously, the above inequality \eqn{m1} is saturated   when
$\Sigma^{(1)}$ is a straight line in the torus. (We have no  
constraints
for the configuration of  $A$.)
Hence  the  desired  BPS mass is obtained as
\ba
{\cal M}_1 = \dil |G|^{-1/4} \sqrt{ q^i G_{ij} q^j }\,\,\,.
\ea

In order to clarify the meaning of the winding $q^i$ in \eqn{m1},
we take a look at the
Wess-Zumino term. The local coordinate of $1$-brane
${\bf R}\times\Sigma^{(1)}$ is expressed by $(t,\sigma)$ and the
Wess-Zumino term turns into a form,
\ba
\int_{{\bf R}\times\Sigma^{(1)}}e^{i\, {\cal F}}\wedge C
&=&\int_{{\bf R}\times\Sigma^{(1)}}
\left(C^{(2)}+ i {\cal F}C^{(0)}\right)\nom\\
&=&q^i\cdot \left(\int_{\bf  
R}\hat{C}^{(2)}_{i0}dt\right)\,\,\,,\nom\\
\hat{C}^{(2)}&:=&C^{(2)}+ i B\,C^{(0)}\,\,\,.    
\label{chat2}
\ea
The $q^i$ turns out to be a ``charge'' associated with the gauge field
$\hat{C}^{(2)}_{i0}$.

~

\noindent
\underline{\em{3-brane}}

Next we consider the $3$-brane case. Let
$(t,\sigma^1,\sigma^2,\sigma^3)$ be a local coordinate of $3$-brane
${{\bf R}\times\Sigma^{(3)}}$. Then one can read the ``charges''  
coupled to
gauge fields $\hat{C}$ 
from the Wess-Zumino term   
\eqn{WZ},
\ba
&&\int_{{\bf R}\times\Sigma^{(3)}}e^{i\, {\cal F}}\wedge  
C\nom\\
&=&\int_{{\bf R}\times\Sigma^{(3)}}\hat{C}^{(4)}
+\int_{{\bf R}\times\Sigma^{(3)}}\frac{i}{2\pi}F\wedge \hat{C}^{(2)}
+\int_{{\bf R}\times\Sigma^{(3)}}\frac{-1}{8\pi^2}F\wedge
   F\cdot C^{(0)}\nom\\
&=& W_l\cdot \int_{\bf R}\frac{1}{6}\,
\epsilon^{lijk}\,\hat{C}^{(4)}_{ijk0}\,dt
+w^i\cdot \int_{\bf R}\hat{C}^{(2)}_{i0}dt\,\,\,,\nom\\
&&\left\{
\begin{array}{ccl}
\hat{C}^{(4)} & := & C^{(4)}+i B\wedge
C^{(2)}-\frac{1}{2}B\wedge B\cdot C^{(0)}\,\,\,,\\
\hat{C}^{(2)} & := & C^{(2)}+i B\cdot
C^{(0)}\,\,\,,
\end{array}
\right.\nom\\
&&\left\{
\begin{array}{ccl}
 W_l & := & {\dis \int_{\Sigma^{(3)}}d^3 \sigma \,\epsilon_{lijk} \,
\frac{\del X^i}{\del \sigma^1}\cdot
\frac{\del X^j}{\del \sigma^2}\cdot
\frac{\del X^k}{\del \sigma^3}}\,\,\,,\\
w^i & := & {\dis \int_{\Sigma^{(3)}}d^3\sigma\,
\epsilon^{\alpha\beta\gamma}
\frac{i}{2\pi}F_{\alpha\beta}\del_{\gamma}X^i}\,\,\,.
\end{array}
\right.
\ea
The $W_l$ is the $3$-brane winding number.
We remark that the non-vanishing expectation value
for $F$ makes the situation a little complicated.
The Poincar{\'e} dual of 
$\dsp \lb \frac{i}{2\pi}F\rb \in H^2(\Sigma^{(3)} ; \bz)$
is represented by  some  1-branes  within our 3-brane  
$\Sigma^{(3)}$ (the case ``branes within branes''  \cite{D}).
The $w^i$ is regarded as the ``effective winding number''
of induced 1-branes.
In this way we can describe by DBI action some  bound states among $3$-branes
and $1$-branes effectively.

Even after this change of situation we can use
the similar  argument for the $1$-brane case
to derive the $3$-brane mass bound,
\ba
{\cal M}_3&=& \dil  |G|^{-1/4}
    \int_{\Sigma^{(3)}} d^3\sigma \,
\sqrt{\det  
\left(G_{\alpha\beta}+B_{\alpha\beta}+F_{\alpha\beta}\right)}   
\nom\\
&\geq &  \dil \sqrt{
(
\begin{array}{cc}
W & w 
\end{array}
)
( M^{IIB,T^4}+J)
\left(
\begin{array}{c}
W\\
w
\end{array}
\right)   }    \,\,\,,    \label{m3}
\ea
where we set
\ba
(\ast  
B)^{ij}&:=&\frac{1}{2\sqrt{|G|}}\epsilon^{ijkl}B_{kl}\,\,\,,\label{*B} 
\\
M^{IIB,T^4}&:=&
\left(
\begin{array}{cc}
\sqrt{|G|}\cdot {\left(G^{-1}-\ast B\cdot G \cdot \ast  
B\right)^{ij}} &
{\left(\ast B\cdot G\right)^i}_j \\
-{\left(G\cdot \ast B\right)_i}^j & {\dis  
\frac{1}{\sqrt{|G|}}G_{ij}}
\end{array}
\right) \,\,\, . 
\label{3-brane moduli}
\ea
It is easy to prove that $8\times 8$ matrix $M^{IIB,T^4}$
belongs to $O(4,4)$ and is also symmetric.
The inequality \eqn{m3}  is saturated when $\Sigma^{(3)}$ is  
a 3-dimensional torus
``linearly embedded'' in $T^4$, that is, the map 
     $X\, :\Sigma^{(3)}\, \rightarrow \, T^4$ is a homomorphism of  
     abelian  groups (not necessarily injective.)
and the gauge field $A$ has a constant  
curvature (whose value is uniquely determined by $w^i$).  
Hence the  desired  mass formula   can be written as\footnote
               {In the situation here, the ``intersection term''
                $(W ~w)J (W ~w)^t$ is always 0.  
               Nevertheless we  wrote this term in \eqn{mass IIB} 
               in order to represent the equivalence with the mass
               formula \eqn{Umass2} manifestly. In the next section
               we will face to the cases 
               that this intersection term essentially appears.}
\be
{\cal M}_3= \dil \sqrt{
(
\begin{array}{cc}
W & w
\end{array}
)
(M^{IIB,T^4} +J)
\left(
\begin{array}{c}
W\\
w
\end{array}
\right)}   \,  \,\,.
\label{mass IIB}
\ee
This formula \eqn{mass IIB} coincides with our algebraic formula \eqn{Umass2}!
Actually the ``moduli matrix'' $M^{IIB,T^4}$ induced from DBI action
is equal to the moduli matrix  $R_s(M)$ derived from U-duality.

To close this subsection we make a remark:
The above description of the systems such that the 3-branes and  
1-branes coexist, i.e.
the bound states of 3- and 1-branes, is not complete.  
For example,
consider the system of the 3-brane wrapping around the 3-torus along
the 678'th axes and 1-brane
wrapping around the circle along the 9'th axis
when the compactified directions are  
the 6789'th axes.
In this situation  one cannot
reinterpret the 1-brane charge as the field strength $F$ as above,  
and cannot derive the correct BPS mass from only the DBI action.
The most naive approach for this problem is to start with the simple  
ansatz for the effective
action
\be
S = S_{\msc{1-brane}}  + S_{\msc{3-brane}} .
\ee
Of course this is not correct, since one must also evaluate
the effect of  interaction  between  these branes by taking the  
sectors
of DN (or ND) open string  into account.  In other words, one must   
calculate
higher loop corrections  to the effective action.
It may be  natural to expect that these loop corrections explain the  
binding energy
 fitted to the conjecture of U-duality.  We will argue on this  
problem in section 4.

~


\subsection{Type IIA on $T^4$}\label{2a}

Next  we analyze the mass formulae  for the type IIA case.
In the same way as in the type IIB case,
we start from the D-brane action \eqn{DBI},  \eqn{WZ}.
The only difference from the type IIB case is that we have
the R-R fields with odd degrees; $C^{(p)}$, $(p=1,3,5,7,9)$,
which leads to the {\em even} D-branes.
In our analysis we need to treat the three kinds of  D-branes  
($0$-,$2$-,$4$-branes).
We first consider the cases that  only the same kind of branes  
exist,
and later discuss  the problem of  the bound states of
branes having different dimensions.

~

\noindent
\underline{\em{0-brane}}

The $0$-brane case is trivial.  The moduli dependence of  BPS mass
only originates from the volume of the internal torus \eqn{dilaton factor};
\be
{\cal M}_0 =\dil |G|^{-1/4} \,  n .
\label{m0}
\ee
The $n$ is the number of $0$-branes and is identified with the  
RR charge
for $C^{(1)}_{\mu}$.

~

\noindent
\underline{\em{2-brane}}
 
For the $2$-brane case,  we take the configuration
$\br \times \Sigma^{(2)}$. As in the analysis of the type IIB case,  
$\br$ is the time axis
and $\Sigma^{(2)}$ is wrapped around some 2-cycle of $T^4$.
Simple evaluation gives the following inequality (which is  
essentially the Minkowski
inequality);
\ba
{\cal M}_2 &=&  \dil |G|^{-1/4} \, \int_{\Sigma^{(2)}}   d^2\sigma \,
           \sqrt{\det  \left(G_{\alpha\beta}   + \cF_{\alpha\beta}  
\right)  }    \nom \\
       &\geq& \dil  |G|^{-1/4} \, \sqrt{  \sum_{i=1}^3\,  
\left(\int_{\Sigma^{(2)}}X^* J_i\right)^2
              + \left( \int_{\Sigma^{(2)}} X^* B \right)^2 }     \,  
\, \, , \label{m2}
\ea
where  $J_1,J_2,J_3\in {H}^{2+} (T^4 ; {\bf R} )$ denotes an  
orthonormal basis
for the self-dual part of $H^2(T^4 ; \br)$ with a normalization
$\dsp J_i \cdot J_j  (\equiv  \int_{T^4} \, J_i \wedge J_j)  =2 |G|^{1/2}  
\delta_{ij}$.
The bound of this inequality \eqn{m2} does not depend on the choice  
of this basis.
Here we assume that $\dsp \int_{\Sigma^{(2)}} \, \cF =  
\int_{\Sigma^{(2)}}\,  X^* B$. It is the case that
 the $U(1)$-gauge field $A$ has no ``monopole'' charge.

When is this inequality \eqn{m2} saturated? 
It is satisfied if the collective coordinate $X$ 
is a holomorphic mapping from  
$\Sigma^{(2)}$ onto
some 2-cycle $S$ determined by a given $\hat{C}^{(3)}_{\mu ij}$ charge.
Strictly speaking, the minimality of  world-volume
does not necessarily mean  a  holomorphic mapping, rather means a   
more
general harmonic mapping. But it is known \cite{BBS}\cite{BVS2} that  
the BPS
condition (the condition of SUSY 2-cycle) leads to a holomorphic  
mapping
at least in the case of $B_{ij}=0$.  We further comment on the  
following fact:
Fix an arbitrary 2-cycle $S \in H_2 (T^4; \bz)$. The condition that   
$S$ is represented
by a holomorphic curve in $T^4$  is that the Poincar{\'e} dual  
$\alpha_S$ of $S$
belongs to $H^{1,1}(T^4; \br)$.  This can be always satisfied  if   
we properly
choose the holomorphic structure of $T^4$.
(The choice of holomorphic structure compatible with
the given metric $G_{ij}$  is parametrized over  $\dsp O(4)/U(2)  
\cong S^2$,
which is identified with $S^2$ spanned by $J_1,J_2,J_3$.   These
degrees of freedom are just equal to 
those needed to make $\alpha_S$  
a (1,1)-form for an arbitrary $S$.)\footnote
       {For the $K3$-compactification we face to the same situation.  
We can still use
the $S^2$ degrees of freedom  for the choice of holomorphic  
structure compatible 
with the Ricci flat metric in order to make the given 2-cycle  
to be algebraic.  On the other hand, a case  
of Calabi-Yau 3-fold forms a striking contrast with it.  In this case  
all the 2-cycles are algebraic from the beginning.}
So, it is sufficient to take $X$ to be a holomorphic mapping  from  
$\Sigma^{(2)}$
onto some holomorphic curve $S$ in $T^4$ with respect to a properly chosen
complex structure.
If we write the corresponding K\"{a}hler form on  
$T^4$ as
$J_S$ normalized by $J_S \cdot J_S =2|G|^{1/2}$,
which is a proper linear combination of $J_1$, $J_2$, $J_3$,
we obtain
\be
    \sqrt{  \sum_{i=1}^3\, \left(\int_{\Sigma^{(2)}}X^*  
J_i\right)^2}
        =  \int_{\Sigma^{(2)}}\, X^* J_S \,\,.
\label{js}
\ee

For the $U(1)$-gauge field $A$, the condition for the saturation  is
somewhat non-trivial.  This is because the pull-back  $X^* B$ is   
not
a harmonic 2-form  even if   all  of the components $B_{ij}$  are  
constants on $T^4$.
Nevertheless we can choose $A$ so that
$\cF (\equiv X^* B + dA) = a X^* J_S$, where $a$ is 
not a function
but merely a complex  
number. (It is most easily proved by making use of the Hodge  
decomposition.)
The assumption $\dsp \int_{\Sigma^{(2)}} \, \cF =  
\int_{\Sigma^{(2)}}\,  X^* B$
immediately gives a relation;
\be
a= \frac{ \int_{\Sigma^{(2)}}\,  X^* B}{ \int_{\Sigma^{(2)}}\,  X^*  
J_S}
   =\frac{ \int_{\Sigma^{(2)}}\,  X^* B}{  \sqrt{  \sum_{i=1}^3\,  
\left(\int_{\Sigma^{(2)}}X^* J_i\right)^2}  } \,\,\, .
\label{a}
\ee
Hence, under this configuration of $X$ and $A$,
we obtain
\ba
{\cal M}_2 &=&  \dil |G|^{-1/4} \, \int_{\Sigma^{(2)}}   d^2\sigma \,
           \sqrt{\det  \left(G_{\alpha\beta}   + \cF_{\alpha\beta}  
\right)  }    \nom \\
&=&  \dil  |G|^{-1/4} \, \sqrt{1+a^2} \, \int_{\Sigma^{(2)}}\, X^* J_S     
   \nom \\
&= &  \dil |G|^{-1/4} \, \sqrt{  \sum_{i=1}^3\,  
\left(\int_{\Sigma^{(2)}}X^* J_i\right)^2
              +  \left( \int_{\Sigma^{(2)}} X^* B \right)^2 }     \,  
\, \, . \label{m22}
\ea
This means the saturation of the inequality \eqn{m2} and
gives the desired  BPS mass formula.

In the similar manner to the case of type IIB 3-brane, one
may consider some extra monopole charge $n$ for $A$
by taking the the background
$\dsp \int_{\Sigma^{(2)}} \, \cF = \int_{\Sigma^{(2)}} \, X^* B  \,  
+ \, n $.
By analyzing the Wess-Zumino term,  we can find that this charge
$n$ can be identified with the extra contribution from the $n$ 0-branes  
within our 2-brane $\Sigma^{(2)}$.  The BPS mass formula can be
easily generalized to this case;
\be
{\cal M}_2 = \dil |G|^{-1/4} \, \sqrt{  \sum_{i=1}^3\,  
\left(\int_{\Sigma^{(2)}}X^* J_i\right)^2
+ \left( \int_{\Sigma^{(2)}} X^* B \right)^2  +
2n \left(\int_{\Sigma^{(2)}}  X^*B\right)+ n^2}   \,\,\,  .
\label{m23}
\ee
This is  indeed  the correct mass formula of the bound states of  a  
2-brane and 0-branes
predicted by U-duality.

~

\noindent
\underline{\em{4-brane}}

In the 4-brane case,
the space-part of world-brane $\Sigma^{(4)}$ must occupy the full  
volume of  $T^4$.
Consider the smooth  map $X : \Sigma^{(4)} \rightarrow  T^4$
covering $T^4$ $m$-times.
We again assume  $\lb \cF \rb = \lb  X^* B \rb$  in  
$H^2(\Sigma^{(4)};\br)$
for the time being.
The inequality of mass integral is represented;
\ba
{\cal M}_4 &=&  \dil |G|^{-1/4} \, \int_{\Sigma^{(4)}}   d^4\sigma \,
           \sqrt{\det  \left(G_{\alpha\beta}   + \cF_{\alpha\beta}  
\right)  }     \nom \\
&\geq & m\,\dil   
\sqrt{ |G|^{1/2}+ 
\left(\int_{T^4} \,B\wedge \ast  B \right)+\frac{1}{4|G|^{1/2}} \left(  
\int_{T^4} B\wedge B\right)^2 }  \,\,\, .
\label{m4}
\ea

What is the condition for saturation?
Clearly the value of this integral does not depend on the choice of  
smooth
map $X$ (under fixing $\mbox{deg}\,  X =m$, of course).
However, the condition for the gauge field  $A$ is
similar to the 2-brane case, but rather complicated,
because the field $X^*B$ necessarily
has constant components.

It reads;
 \be
\cF\wedge\cF = b \, dv \, , ~~~
 \cF \wedge J_i    = b_i \, dv \,\,\,  (i=1,2,3)\,\,\,,
\label{condition 4-brane}
\ee
for some complex numbers $b,~b_i$, and
$dv$ denotes the volume form of $T^4$.
The values of  these constants can be easily solved in the same way  
as
\eqn{a} and we obtain the BPS mass formula;
\be
{\cal M}_4 = m\, \dil 
\sqrt{ |G|^{1/2}
+ \left(\int_{T^4} \,B\wedge \ast  B  \right) +  \frac{1}{4|G|^{1/2}} 
\left(  \int_{T^4} B\wedge B\right)^2 }  \,\,\, .
\label{m42}
\ee

Now let us consider the case that the gauge field $A$ has a  
non-trivial
topological charge. Alternatively one may regard it as an ``integer  
theta-parameter shift'' \cite{GPR},  
$B_{ij}  \rightarrow  B_{ij} + \Theta_{ij}$   
$({}^\forall \Theta \in H^2(T^4; \bz))$,
which is a part of T-duality transformations. 
By analyzing the Wess-Zumino term again, we can find that
$\dsp \lb \frac{i}{2\pi}F \rb \in H^2(T^4;\bz)$ 
is identified with the extra 2-brane  charge
and $\dsp \lb -\frac{1}{8\pi^2}  F\wedge F\rb \in
 H^4(T^4;\bz) (\cong \bz)$ is identified  
with the extra 0-brane charge.  
The corresponding mass formula is immediately  calculated.
This result is rather complicated, but   we can show that 
it is fitted to the similar formula  to \eqn{mass IIB};
\be
{\cal  M}_4  = \dil \sqrt{  Q^T\, (M^{IIA,T^4}  +J)  \,  Q}  \,\, ,
\label{mass IIA}
\ee
where the ``charge vector''
$Q\in \bz^{\oplus 8}$  describes the RR charge  of  4-branes ($\cong
{\bz}$) and the effectively induced 2-brane ($\cong {\bz}^{\oplus 6}$)
and 0-brane ($\cong {\bz}$) charges. 
$M^{IIA,T^4} $  is also a symmetric matrix valued in $O(4,4)$.  Here we do  
not present  its explicit form, since it is cumbersome compared with  
the type IIB case.

 We only point out the following:
Firstly, all of three cases (the 0-brane case \eqn{m0},
2-brane  cases \eqn{m22} and the cases of bound states of 0-branes  
and 2-branes
\eqn{m23}) are also summarized by the same formula \eqn{mass IIA}.
We only have to set the 4-brane charge to be zero in the 
vector $Q$.

Secondly,  in the similar manner to the type IIB case,
the moduli matrix  $M^{IIA,T^4} $ can be identified with 
the cospinor representation matrix $ R_c(M) $ of 
$O(4,4)$. 
This means the consistency of our mass formulae \eqn{mass IIA}
with the prediction of U-duality.
This result  is also expected from the fact that the replacement  
$M^{IIA,T^4} \,\leftrightarrow \,M^{IIB,T^4}$ by T-duality should occur.
But the check of this claim is not so self-evident,
because the moduli matrices become rather complicated under general 
background. In this sense,  we may also say our results support 
the consistency of DBI action under T-duality.   

Lastly, we comment  on the same difficulty as that 
in the analysis of type IIB case.  One must understand that
generic bound states of 0-, 2-, and 4-branes cannot be described by  
the recipe of  the branes within branes \cite{D}.  We can at most   
analyze, by using  only the DBI action,   the cases
which can be connected by the T-duality transformations\footnote
     {Here we only consider the charges of ``$U(1)$-sectors'' of 
       gauge fields. We will present in the next section
       some comments on the possibility
       to develop our analysis to that based 
       on the ``non-abelian Born-Infeld action (NBI)'' 
       proposed by Tseytlin \cite{NBI}.}
$B_{ij} ~ \rightarrow ~ B_{ij} + \Theta_{ij}$   with the cases   
that  4-branes alone exist.
Of course, one cannot describe all the bound states  by making use  
of these T-duality transformations.  In order to complete   
our analyses  we need to argue  seriously on  the binding energies 
of branes.
In section 4 we will return to this problem.

~
\newpage
\subsection{Considerations from M-theory}\label{2}

In this subsection we reconsider the case of $IIA$ on $T^4$ by the M-theory
approach. The discussion here is based on that of \cite{susyDVV}.

The M-theory is believed to exist in $11$ dimensional space-time,
whose massless multiplet is a $11$ dim. supergravity one,
$\hat{g}_{MN}$, $c_{KLM}$, and $\psi^{\alpha}_M$.
Here the $\hat{g}_{MN}$ is the $11$ dimensional graviton and $c_{KLM}$
is the rank $3$ anti-symmetric field. The spinor index ``$\alpha$'' of
the gravitino $\psi^{\alpha}_{M}$ labels a $32$ component Majorana
fermion. The indices ``$K,L,M, \cdots$'' represent space-time
coordinates and run from $0$ to $10$.

Let us consider a compactification of this theory on the
$5$-dimensional torus $T^5$. The $6$-dimensional space-time
is extended in the $({X^0},{X^1},\cdots ,{X^5})$-directions
and the remaining coordinates $(X^m)$ $(m=6,7,8,9,10)$
represent the torus $T^5$.
Especially the $5$th coordinate ``$X^{10}$'' of the $T^5$ is the
``longitudinal'' one in this $11$ dimensional theory.
The $T^5$ is reexpressed as a product $T^5 = T^4 \times S^1 $
of a transverse torus $T^4$ with $(X^{\bar{m}})$ $(\bar{m}=6,7,8,9)$
and a circle $S^1$ parametrized by the $X^{10}$.
(We use indices ``$\mu ,\nu ,\lambda ,\cdots$'' for the $6$ dimensional
space time.)
This compactified theory M/$T^5$ on $T^5 (=T^4 \times S^1)$ is
equivalent to the IIA string on $T^4$ because of the string duality
``$\mbox{M/}S^1 \cong \mbox{IIA}$''.

The moduli space of this is expressed by 
a $6$ dimensional dilaton $\Phi$ and
$25$ scalars including $\hat{g}_{mn}$'s and $c_{lmn}$'s.
These $25$ scalars parametrize the homogeneous space 
$O(5,5)/(O(5)\times O(5))$ (U-duality group).
In the context of the IIA string, the NS-NS scalar part contains $17$
scalars ({\small $i.e.$} $10$ $\hat{g}_{\bar{m}\bar{n}}$, $6$ 
$B_{\bar{m}\bar{n}}= c_{\bar{m}\bar{n} 10}$ and a
$\hat{g}_{10,10}$). The remaining eight scalars, {\small $i.e.$} 
$4$ $C^{(1)}_{\bar{m}}=\hat{g}_{\bar{m}10}$
and $4$ $C^{(3)}_{\bar{l}\bar{m}\bar{n}}=c_{\bar{l}\bar{m}\bar{n}}$, 
are combined into a R-R
moduli part.

The BPS states are characterised by their charges. There are $16$
charges in the M/$T^5$-theory. Vectors $\hat{g}_{{\mu}{m}}$ couple to a charge
vector $r_m$ and $c_{\mu m n }$ and its dual $c^{*}_{\mu
klmnr}$ similarly couple to $10$ charges $s^{mn}$  
and one charge $q$ respectively.
From the point of view of the string theory (IIA/$T^4$-theory),
the momenta $p_{\bar{m}}$, winding numbers $w^{\bar{n}}$ in the NS
sector are identified with $r_{\bar{m}}$, $ s^{\bar{n} 10}$
respectively. Also we can identify D0-, D4-brane charges $q_4$, $q_0$
with $q_4 =q$, $q_0 = - r_{10}$ and six D2-brane
charges $s^{\bar{m}\bar{n}}$ are combined into a rank $2$
anti-symmetric form 
$q_2 :=\frac{1}{4}\epsilon_{\bar{k}\bar{l}\bar{m}\bar{n}}
  s^{\bar{k}\bar{l}} d X^{\bar{m}}\wedge d X^{\bar{n}}$. 
It is useful to note that $q_2$ is the Poincar\'{e} dual of 
the homology cycle around which the D2-brane is wrapping; 
$\Sigma := s^{\bar{k}\bar{l}}\Sigma_{\bar{k}\bar{l}}$, 
where $\Sigma_{\bar{k}\bar{l}}$ $(\bar{k} < \bar{l})$ denote 
the bases of $H_2(T^4)$ defined by the relation
$ \dsp
\int_{\Sigma_{\bar{k}\bar{l}}} \, d X^{\bar{m}}\wedge d X^{\bar{n}}=
  \delta^{\bar{m}}_{\lb \bar{k} }\, \delta^{\bar{n}}_{\bar{l}\rb }.
$

We summarize 
these scalars, vectors and
charges
of the M/$T^5$-theory (equivalently IIA/$T^4$-theory) in the tables 
\ref{zu2},\ref{zu3}. 

\newpage

\begin{table}[h]
\begin{center}
\begin{tabular}{|ll|ll|ll|}\hline
 & & \multicolumn{4}{c|}{IIA/$T^4$}\\
\cline{3-6}
\multicolumn{2}{|c|}{\raisebox{1.5ex}[0pt]{M/$T^5$}}
 &  NS-NS & (16) & R-R & (8) \\\hline
$\hat{g}_{mn}$ & (15) & $\hat{g}_{\bar{m}\bar{n}}$ & (10) &
$ C^{(1)}_{\bar{m}}= \hat{g}_{\bar{m}10} $ & (4) \\
$c_{lmn}$ & (10) &
$B_{\bar{m}\bar{n}}= c_{\bar{m}\bar{n}10}$ 
& (6) &
$C^{(3)}_{\bar{l}\bar{m}\bar{n}}$ & (4) \\
$\Phi $ & (1) &
$\hat{g}_{10,10}$ & (1) & &  \\\hline
\end{tabular}
\caption{Scalar fields of M/$T^5$ and IIA/$T^4$ \label{zu2}}
\end{center}
{\small The M/$T^5$ theory has scalars $\hat{g}_{mn}$, $c_{lmn}$ derived from
the graviton, rank $3$ anti-symmetric field in 11 dimensional SUGRA.
The $\Phi$ is a six dimensional dilaton combined with the volume of
$T^5$ and 10 dimensional dilaton $\phi $.
In NS-NS sector of the IIA/$T^4$ side, 
$\hat{g}_{\bar{m}\bar{n}}$, $B_{\bar{m}\bar{n}}$
are scalar fields and $\hat{g}_{10,10}$ is essentially ten dimensional dilaton.
There are RR $1$-form $C^{(1)}$ and RR $3$-form $C^{(3)}$
in the RR-sector. Also
numbers ``$\sharp$'' in parentheses $(\sharp)$ are degrees of freedom of
associated scalars.}
\end{table}


\begin{table}[h]
\begin{center}
\begin{tabular}{|ll|ll|ll|}\hline
 & & \multicolumn{4}{c|}{IIA/$T^4$}\\
\cline{3-6}
\multicolumn{2}{|c|}{\raisebox{1.5ex}[0pt]{M/$T^5$}}
 &  NS-NS & (8) & R-R & (8) \\\hline
$\hat{g}_{\mu m}$ & ($r_{m}$ ; 5) & $\hat{g}_{\mu \bar{m}}$ & ($r_{\bar{m}}$ ; 4) &
$ C^{(1)}_{\mu}= \hat{g}_{\mu 10} $ & ($r_{10}$ ; 1) \\
$c_{\mu mn}$ & ($s^{mn}$ ; 10) &
$B_{\mu \bar{m}}= c_{\mu\bar{m}10}$ 
& ($s^{\bar{m}10}$ ; 4) &
$C^{(3)}_{\mu \bar{m}\bar{n}}$ & ($s^{\bar{m}\bar{n}}$ ; 6) \\
$c^{\ast}_{\mu klmnr} $ & ( $q$ ; 1) &
 &  & $C^{(6)}_{\mu \bar{k}\bar{l}\bar{m}\bar{n}}$ & ($q$ ; 1) \\\hline
\end{tabular}
\caption{Charges and Vector fields of M/$T^5$ and IIA/$T^4$ \label{zu3}}
\end{center}
{\small There are three kinds of vector fields $\hat{g}_{\mu m}$,
$c_{\mu mn}$, and its dual $c^{\ast}_{\mu klmnr}$ 
in the M/$T^5$ theory. They can couple to charges $r_m$, $s^{mn}$ and $q$
respectively. 
In the context of the IIA string, eight vectors $\hat{g}_{\mu\bar{m}}$,
$c_{\mu\bar{m}10}$ are combined into a NS sector multiplet.
Their associated charges $r_{\bar{m}}$, $s^{\bar{m}10}$
are called momenta $p_{\bar{m}}$, winding numbers $w^{\bar{m}}$
respectively.
In the R-sector, the vector multiplet consists of
the remaining eight vectors $\hat{g}_{\mu 10}$,
$C^{(3)}$ and $C^{(6)}$.
Numbers of parentheses represent the degrees of freedom of the
corresponding charges.}
\end{table}

\newpage


Now let us consider BPS mass formulae from supersymmetric algebra.
Firstly the resulting theory $\mbox{M/}T^5 =\mbox{IIA/}T^4$ has
a space-time $6$-dimensional $(2,2)$ supersymmetry.\footnote
    {This 6-dimensional SUSY is often called ``(4,4)'' in some literature.} 
The SUSY algebra
has an internal $SO(5)\cong USp(4)$ R-symmetry and there is a
symplectic structure $\omega_{ab}(=-\omega_{ba})$ $(a,b=1,2,3,4)$,
and we denote $\omega^{ab}:= \omega^{-1}_{ab}(\equiv -\omega_{ab})$.
Super charges $Q^a_{\alpha}$, $\bar{Q}^b_{\bar{\beta}}$ are
$4$-component $USp(4)$-(pseudo)Majorana Weyl spinors. The Latin indices
``$a,b$'' label the R-symmetry $USp(4)$. 
The Greek indices ``$\alpha
,\bar{\beta} $'' $(\alpha ,\bar{\beta} =1,2,3,4)$ 
represent $4$-component Weyl spinors in $6$ dimension.
The Latin indices should be raised and lowered by the 
``charge conjugation matrices'' $\omega^{ab}$ and $\omega_{ab}$. 
For example, $\omega_{ab}Q^b_{\alpha}= Q_{a\alpha}$.

These charges satisfy the following  SUSY algebra
\begin{eqnarray}
&&\{ Q^a_{\alpha} ,Q^b_{\beta} \}=
\omega^{ab} {\gamma^{\mu}}_{\alpha\beta} p_{\mu}\,\,\,,\nonumber \\
&&\{ Q^a_{\alpha} ,\bar{Q}^b_{\bar{\beta}} \}=
\delta_{\alpha \bar{\beta}} Z^{ab}
\,\,\,,\nonumber 
\end{eqnarray}
where ${\gamma^{\mu}}_{\alpha \beta}$'s are $6$ dim. space-time
$\gamma$-matrices and $p_{\mu}$ is the space-time momentum.
The central charge $Z^a_{~b}$ can be decomposed by internal $SO(5)$ 
$\gamma$-matrices ${\Gamma^m}_{ab}$ $(m=6,7,8,9,10)$,
\begin{eqnarray}
&& Z^a_{~b}=\left( q \sqrt{\det \hat{g}}\,\, {\bf 1}
    + r_m \Gamma^m +\frac{1}{2} s^{mn}
   \Gamma_{mn}\right)^a_{~b} \,\,\,,\\
&& \Gamma_{mn}:=\Gamma_{[m }\Gamma_{ n] }  
  \,\,,\,\,\,
\{\Gamma_m ,\Gamma_n \}=2 \hat{g}_{mn} \,\,\,.\nonumber
\end{eqnarray}
When we turn off the Kalb-Ramond field $B_{\bar{m}\bar{n}}$ and RR
fields $C^{(1)}$, $C^{(3)}$ for simplicity, the coefficients 
$q$, $r_m$, $s^{mn}$ of the decomposition can be identified with the
previous charges of the M/$T^5$-theory. 
In the following, we abbreviate space-time spinor indices $\alpha$, 
$\bar{\beta}$.

The $1/4$-susy condition of BPS states can be written
\begin{eqnarray}
(\epsilon_a Q^a +\bar{\epsilon}_b \bar{Q}^b) \left| BPS \right\rangle
=0.\nonumber
\end{eqnarray}
This condition is transformed into eigenvalue problems of
operators $Z Z^{\dagger}$, $Z^{\dagger} Z$ with eigenvectors
$\epsilon$,
$\bar{\epsilon}$
\begin{eqnarray}
&&{(Z Z^{\dagger})^a}_b {\epsilon^b} 
=\{{( m_0^2 {\bf 1} +2 (K_m + W_m ) \Gamma^m)^a}_b\} {\epsilon^b} 
= m_{BPS}^2 {\epsilon^a} \,\,\,,\nonumber \\
&&{( Z^{\dagger} Z)_a}^b {\bar{\epsilon}_b} 
=\{{( m_0^2 {\bf 1} +2 (K_m - W_m ) \Gamma^m )_a}^b\} {\bar{\epsilon}_b} 
= m_{BPS}^2 {\bar{\epsilon}_a }\,\,\,,\nonumber \\
&& m_0^2 :=q^2 \det \hat{g} +\hat{g}^{mn} r_m r_n +\frac{1}{2} \hat{g}_{lm}
\hat{g}_{kn} 
s^{lk} s^{mn} \,\,\,,\nonumber \\
&& K_s :=  \left( q r_s -\frac{1}{8} s^{kl} s^{mn} 
   \epsilon_{klmns} \right)\sqrt{\det \hat{g}}
\,\,\,,  \label{M0} \\
&& W^s := r_l  s^{ls} \,\,\,,\nonumber \\
&& \,\,\,\,\,\, Z^{\dagger} := \omega Z \omega^T \,\,\,,\,\,\,
\omega^T = -\omega \,\,\,. \nonumber
\end{eqnarray}
Here the 
$m_{BPS}^2 :=- p^2$ is the square of BPS mass and 
is given explicitly 
\begin{eqnarray}
&& m_{BPS}^2 = m_0^2 \pm 2 \sqrt{\hat{g}_{mn} W^m W^n +\hat{g}^{mn} K_m K_n}
\,\,\,.\label{susyMass}
\end{eqnarray}
The ambiguity of the signature in Eq.(\ref{susyMass})
depends on whether the state is a BPS or anti-BPS state.
In the following we restrict calculations to the BPS case and choose
the ``$+$'' sign in Eq.(\ref{susyMass}).

Next we will rewrite this BPS mass formula in the IIA/$T^4$ side.
We have switched off all the RR scalars
$\hat{g}_{\bar{l}10}=C^{(1)}_{\bar{l}}=0$ (RR 1-form),
and $C^{(3)}_{\bar{k}\bar{l}\bar{m}}=0$ (RR 3-form)
and the Kalb-Ramond field 
$B_{\bar{k}\bar{l}} = c_{\bar{k}\bar{l}10}$.
The charges in the NS sector are momenta $p_{\bar{m}}(\equiv r_{\bar{m}})$ and
winding numbers  $w^{\bar{m}}(\equiv s^{\bar{m} 10})$.
On the other hand, remember that 
the RR-sector charges are classified into three types;
$q_4 \equiv q$ (4-brane charge),
$q_2 \equiv 
\frac{1}{4}\epsilon_{\bar{k}\bar{l}\bar{m}\bar{n}}
  s^{\bar{k}\bar{l}} d X^{\bar{m}}\wedge d X^{\bar{n}}$ 
(2-brane charge), and
$q_0 \equiv - r_{10} $ (0-brane charge).

The BPS mass formula (\ref{susyMass}) is evaluated by the Einstein
metric $\hat{g}_{mn}$ in 11 dimension. However, we want to compare this
formula with the results in the previous subsection, and 
it was calculated in the 6 dimensional Einstein frame.
Hence we will write down the relations between the 11D metric
$\hat{g}_{mn}$, 10D string metric $G_{mn}$, and 6D Einstein
metric $g^{(6)}_{mn}$,
\begin{eqnarray}
&&\hat{g}_{10,10}:=e^{\frac{4}{3}\phi}\,\,, \nonumber\\
&&\hat{g}_{\bar{m}\bar{n}}
=e^{-\frac{2}{3}\phi}G_{\bar{m}\bar{n}}\,\,\,\,
(\bar{m},\bar{n}=0,1,\cdots ,9)\,\,,
\nonumber \\
&&G_{\mu\nu}=e^{\Phi}g^{(6)}_{\mu\nu}\,\,\,\,(\mu ,\nu =0,1,\cdots
5)\,\,.\nonumber
\end{eqnarray}
The $\Phi$ is the 6 dimensional dilaton and is associated with the
10 dim. dilaton $\phi$ and the volume of the $T^4$
as
\begin{eqnarray}
&&e^{\Phi}=e^{\phi}|G|^{-1/4}\,\,,\nonumber \\
&&\,\,|G|:=\det (G_{ij})\,\,\,\,\,(i,j =6,7,8,9)\,\,\,.\nonumber
\end{eqnarray}
Note a relation;
\begin{eqnarray}
\sqrt{\hat{g}_{00}}&=&e^{-\frac{1}{3}\phi}\sqrt{{G}_{00}}\nonumber\\
&=&e^{-\frac{1}{3}\phi}e^{\frac{1}{2}\Phi}\sqrt{{g}^{(6)}_{00}}\,\,\,.\nonumber
\end{eqnarray}
Thus we have to multiply a factor
$e^{-\frac{1}{3}\phi}e^{\frac{1}{2}\Phi}$
to results obtained from Eq.(\ref{susyMass}) in order to
evaluate BPS mass $m_{BPS}^{(6)}$ in the 6 dimensional Einstein frame.

As a first case, we concentrate on the mass formula in the NS-NS sector
with no D-brane charges. We can evaluate its BPS mass from the
Eq.(\ref{susyMass})
\begin{eqnarray}
(m^{(6)}_{BPS})^2&=&e^{\Phi}\left(
G^{\bar{m}\bar{n}}p_{\bar{m}}p_{\bar{n}}+
w^{\bar{l}} {G}_{\bar{l}\bar{k}}
w^{\bar{k}}
+ 2p_{\bar{m}} {\delta^{\bar{m}}}_{\bar{k}}
w^{\bar{k}}\right).
\end{eqnarray}
As a second case, the mass formula with only D-brane charges
can be calculated as
\begin{eqnarray}
 (m^{(6)}_{BPS})^2 
&=& e^{-\Phi}|G|^{-1/2}\left[( {q_4}|G|^{1/2} + {q}_0 )^2 +
2 |G|^{1/2} ( {q}_2^{+}\cdot {q}_2^{+ })\right],\label{mBPS1} \\
q_2^{+}&:=&\frac{1+ \ast}{2}q_2 \nonumber ,
\end{eqnarray}
where $\ast $ is a Hodge dual in the $T^4$ and
$q_2^{+}$ 
corresponds to the self-dual part of $q_2$.
For arbitrary 2-forms $A_1$, $A_2$ on $T^4$, the inner product $(A_1
\cdot A_2)$ is defined as
\begin{eqnarray}
(A_1 \cdot A_2):=\int_{T^4}A_1 \wedge A_2\,\,\,.\nonumber
\end{eqnarray}


If we want to incorporate  further 
the Kalb-Ramond moduli $B_{\bar{k}\bar{l}}\not= 0$, 
we only have to perform a suitable $O(5,5)$-rotation on the charge vectors
$q$, $r_l$, $s^{kl}$ (corresponding to the element $\Omega_B$ 
of $O(4,4)$-subgroup in Eq.\eqn{OmegaB}). 
Namely, we should replace $q$, $r_l$, $s^{kl}$
in the expressions of $m_0$, $K_s$
and $W^s$ in Eqs.\eqn{M0} by the following $\tilde{q}$, $\tilde{r}_l$ and 
$\tilde{s}^{kl}$.
\begin{eqnarray}
&&\tilde{q}:= q\,\,, \nonumber 
\\
&&\tilde{r}_l:=r_l+\frac{1}{2} s^{mn}c_{mnl}
  + \frac{1}{4}q\sqrt{\det \hat{g}}\,\,(\ast_5 c)^{mn}c_{mnl}\,\, ,\nonumber \\
&&\tilde{s}^{kl}:=s^{kl} + q\sqrt{\det \hat{g}}\,\,(\ast_5 c)^{kl}\,\, .\nonumber
\end{eqnarray}
Here the symbol ``$\ast_5 $'' 
means the Hodge dual in the $T^5$.
Then the BPS mass formula is expressed 
in the same form as that of Eq.(\ref{susyMass})
with these replacements.

The NS-NS sector BPS mass can be written as
\begin{eqnarray}
(m^{(6)}_{BPS})^2&=&e^{\Phi}\Bigl[
G^{\bar{m}\bar{n}}p_{\bar{m}}p_{\bar{n}}+
w^{\bar{l}}( {G}_{\bar{l}\bar{k}}
-B_{\bar{l}\bar{m}}G^{\bar{m}\bar{n}}B_{\bar{n}\bar{k}})w^{\bar{k}}\nonumber\\
&&+2p_{\bar{m}}( {\delta^{\bar{m}}}_{\bar{k}}
-G^{\bar{m}\bar{n}}B_{\bar{n}\bar{k}})w^{\bar{k}}\Bigr]\,\,\,.
\end{eqnarray}
It reproduces the well-known BPS mass formula of fundamental 
string excitations.
Similarly we obtain RR sector mass formula
\begin{eqnarray}
(m^{(6)}_{BPS})^2 &=&e^{-\Phi} \Biggl[
|G|^{-1/2} q_0^2 +q_4^2 
\left\{|G|^{1/2}+ (B\cdot\ast B)+{\displaystyle \frac{1}{4 |G|^{1/2}}}(B\cdot
B)^2 \right\}\nonumber\\
&&+ |G|^{-1/2} \left\{ (q_{2}\cdot B)^2 + 2 |G|^{1/2}
(q_2^{+}\cdot q_2^{+})\right\}\nonumber \\
&&+q_0 q_4 \left\{2 - |G|^{-1/2}(B\cdot B)\right\}
         - 2 q_0 |G|^{-1/2}(q_{2}\cdot B)\nonumber \\
&&+ q_4 \left\{|G|^{-1/2} (B\cdot B)(q_{2}\cdot B)+
2 (q_2 \cdot * B) \right\}
\Biggr]\,\,\,.
\end{eqnarray}
As special cases when there are only one kind of RR charges,
we write down results;
\begin{eqnarray}
\bullet \,\,&& \,\, \underline{\mbox{0-brane}}\nonumber  \\
&&(m^{(6)}_{BPS})^2=e^{-\Phi} |G|^{-1/2} q_0^2 ,\\
\bullet \,\,&& \,\, \underline{\mbox{2-brane}} \nonumber  \\
&&(m^{(6)}_{BPS})^2
=e^{-\Phi} |G|^{-1/2}
\left\{  (  q_2 \cdot B)^2 
+ 2 |G|^{1/2} (q_2^{+}\cdot  q_2^{+})\right\},\label{D2mass}\\
\bullet \,\,&& \,\, \underline{\mbox{4-brane}}\nonumber\\
&&(m^{(6)}_{BPS})^2=e^{-\Phi}  q_4^2
\left[|G|^{1/2}+ (B\cdot \ast B) +\frac{1}{4}|G|^{-1/2}(B\cdot B)^2 
\right].
\end{eqnarray}
These results can be compared with 
evaluations from D-brane techniques in the previous subsection and
there are completely agreements between them.
(It is useful to remark  
the simple relations $\dsp 2|G|^{1/2}(q_2^{+}\cdot  q_2^{+})
= \sum_{i=1}^3 \, \left(\int_{\Sigma} J_i\right)^2$, 
$\dsp q_2 \cdot B = \int_{\Sigma}B$, where $\Sigma \in H_2(T^4)$ denotes
the Poincar\'{e} dual of $q_2$.)
Also the mass formula of D0-D4 system is written down
\begin{eqnarray}
({m^{(6)}_{BPS}})^2
&=& e^{-\Phi} \Biggl( |G|^{-1/4}q_0 +q_4 \Bigl\{|G|^{1/2}+(B\cdot \ast B)
+{\displaystyle \frac{1}{4 |G|^{1/2}}(B\cdot B)^2}
\Bigr\}^{1/2} \Biggr)^2 \nonumber \\
&&-2 e^{-\Phi} \Biggl[
\Bigl\{1+|G|^{-1/2}(B\cdot \ast B)
+{\displaystyle \frac{1}{4|G|}(B\cdot B)^2} \Bigr\}^{1/2}
-\left(1-{\displaystyle \frac{(B\cdot B)}{2 |G|^{1/2}}}\right)
\Biggr] q_0 q_4 \label{D0D4mass}\,\,\,.\nonumber \\
&&
\end{eqnarray}
These results Eqs.(\ref{D2mass}), (\ref{D0D4mass}) will be compared to
the binding energy calculations in the next section.

~

~

\subsection{Type IIA on $K3$}

The type IIA string compactified on $K3$ is very similar
to the case of type IIA over $T^4$.  This is because  $K3$
has orbifold limits described by $T^4/ \bz_2$.  But one needs to
take account of the extra  64 moduli and 16 gauge fields associated
with the degrees of freedom of blow-ups of  16 fixed points
(which correspond to the 16 matter multiplets in the sense of $6D$  
$N=(1,1)$ theory \cite{Comment} 
and to the twisted sectors in the language of orbifold CFT).

Our above analysis of D-brane masses is also applicable to this  
case.
The calculation is almost parallel to the case of type IIA over  
$T^4$.
We can summarize this  result as follows;
\be
{\cal  M}  = \dil \sqrt{  Q^T\, (M^{IIA,K3}  +L)  \,  Q}  \,\, .
\label{mass IIA K3}
\ee
Here the charge vector $Q$ has  $24 (=  8+16)$ components and
the moduli matrix $M^{IIA,K3} $ is an $O(4,20)$
symmetric matrix which parametrizes the K3-moduli space
$\dsp O(4,20)/(O(4) \times O(20))$ \cite{As}. The ``intersection matrix''
$L$ can be expressed as
$L= (E_8(-1))^{\oplus 2} \, \oplus \sigma_1^{\oplus 4}$ in the standard  
basis.

This formula \eqn{mass IIA K3} is manifestly consistent with the  
famous conjecture of type IIA - heterotic duality in 6-dimension \cite{HT},
since the duality transformation maps the K3-moduli matrix $M^{IIA,K3} $
into the matrix of Narain moduli space $M^{het, T^4}$ of heterotic  
string.

However, there is a crucial difference from the case of $T^4$, which  
is due to the fact that
$K3$ is a curved manifold.  It is known \cite{BVS2,GHM} that the 4-brane  
in the $K3$ case
has the extra 0-brane charge $-1$
due to the 1st Pontrjagin number of $K3$.
This leads to unexpected
assignments of R-R charges to the configurations  of  
D-branes,
and gives a serious contradiction to our analysis of  the BPS mass  
spectrum of the R-R solitons.
In order to get over  this difficulty  we will have to take account of 
the extra degrees of freedom that are absent in the $T^4$-case,
the twisted sectors in $T^4 / \bz_2 \cong  K3$.
If the contributions from the twisted sectors to the effective action
(or the equations of motion) can be interpreted as bound states of 
4-branes and {\em effective\/} 0-branes, which perhaps reside at 
the fixed points of $\bz_2 $-action, the similar analyses to 
those in the next section might yield the correct results.
However, our study on this problem is still far from the complete solution.
We would like to present a further discussion elsewhere.

~

~

\section{Higher-Loop Corrections to the Mass Formula as the Binding Energies 
among Intersecting Branes}
\cleqn

In this  section we shall further develop our D-brane analysis for BPS
mass spectra. 
In our previous analyses in sections 3.1, 3.2 
we only used the 1-loop effective actions of D-branes (DBI action). 
We already mentioned the limits of 1-loop analysis. 
We will face to the problems:
\begin{enumerate}
\item Calculations of 
  the binding energies of the bound states of D-branes which cannot be  
  evaluated from the DBI action.
\item Calculations of 
  the binding energies of the bound states of D-branes and 
  fundamental excitations of string (in the cases of 
   type IIA and type IIB over $T^4$). 
\item Evaluation of the dependence of the BPS masses  on  the R-R moduli 
  (in the cases of type IIA and type IIB over $T^4$).
\item Evaluation of the correction to the mass formula from
  the extra 0-brane charge due to the 1st Pontrjagin number
   (in the case of type IIA over K3).
\end{enumerate} 

These four problems share an important point;
they all may   be resolved   by analyzing $N(\geq 2)-$loop
 corrections to the D-brane actions, or equivalently, 
to the $\beta$-functions. In this section we present a  higher loop
analysis and show that this statement is indeed true for the 1st problem.
The 2nd, 3rd, 4th problems are more challenging, but we believe that 
the higher loop analyses will also give the correct answers. 
We would like to discuss these problems elsewhere.  

To solve the 1st problem,  we will discuss an important relationship 
between the binding energies of the bound states and some 
SUSY breaking. This statement may sound strange, 
since we should now consider the mass spectra of BPS solitons 
which should preserve a part of SUSY! 
But this is not a contradiction. One must carefully understand
the term ``BPS''. 
This should be used in the framework
of the 6-dimensional supergravity theory
which is the low energy effective theory compactified over 
the 4-torus. On the other hand, if we interpret the BPS solitons with 
RR charges as D-branes, we must treat the full
10-dimensional superstring theory.
Of course, the states with some unbroken SUSY in the sense of 10-dimensional 
theory are also supersymmetric in the sense of 6-dimensional effective theory.
But the inverse is {\em not\/} correct. 
Actually, we will later focus on some brane 
configurations which break SUSY in the sense of 10-dimensional string theory
but should correspond to the BPS states in the sense of 6-dimensional SUGRA. 
We can expect that even if  these states have no higher loop corrections 
in the framework of 6-dimensional  SUGRA, they can have {\em stringy\/} 
loop corrections in the framework of 10-dimensional superstring.
In this sense we may say our calculation of binding energies will give 
a non-trivial check of U-duality {\em in the level of quantum
string theory.}

~

\subsection{Higher Loop Corrections to $\beta$-Functions and 
SUSY Breaking}

In the sequel  we consider the type II (A or B) string over $T^4$.
Let us start with the loop corrected equations of motion of string\footnote
    {The ``loop'' means the sum of world sheet genera
    and boundary loops. We evaluate the contributions to the $\beta$ function
     from these world sheet loops. From the point of view of the sigma model
     perturbation, we calculate the quantity in the first order of the
     $\alpha'$-expansion.}.
Throughout this section we take a convention $\mu, \, \nu = 0, \ldots , 5$
(6-dimensional space-time),
$i, \, j = 6,\ldots, 9$ (the internal torus).
We also use the notation $|G|= \det \, G_{ij}$ (the square of volume of 
internal torus).
Recall that 
 $G_{\mu\nu}= e^{\Phi} g_{\mu\nu}^{(6)}\equiv e^{\phi}|G|^{-1/4} 
g_{\mu\nu}^{(6)}$, where $g_{\mu\nu}^{(6)}$ denotes the 6-dimensional
Einstein frame metric.
Set $ g_{\mu\nu}^{(6)} = \eta_{\mu\nu}+ h_{\mu\nu})$. 
The (linearized) equation of motion for $h_{00}$ can be written as 
\be
\beta_{h_{00}}\equiv  - \la^{-2} \mbox{\Large$\Box$} h_{00}(x) 
+ c_{(1)}(x)+ c_{(2)}(x)+
\cdots + c_{(n)}(x) + \cdots = 0 .
\label{beta function}
\ee
Here $\la$ is the string coupling constant $\la \equiv e^{\phi}$
and $c_{(n)}$ denotes the $n$-loop contribution to $\beta_{h_{00}}$
($n:= \sharp \, \{ \mbox{open string loops}\} \, 
+ 2 \sharp \, \{\mbox{closed string loops}\} $).
Of course $c_{(n)}$ should have the $\la$-dependence $\sim \la^{n-2}$.
In our setting, both the $h_{00}(x)$ and $c_{(n)}(x)$ do not depend on 
the coordinates along the internal torus $x^6,\, \ldots , \,x^9$
(and also we assume that they do not depend on the time $x^0$, 
since we are now considering a static problem), so the equation of motion
\eqn{beta function} is reduced to that in the 6-dimensional space-time.

Consider a Dirichlet $p$-brane ${\cal D}$ wrapping  around an internal 
$p$-cycle so as to be  observed as a rest particle 
in our  6-dimensional space-time. 
We can rewrite the equation of motion \eqn{beta function} 
as 
\be
\la^{-2} \mbox{\Large $\Box$} h_{00}(x) = 
\sum_{n\geq 1} \, c_{(n)}(x)\,\,\,.
\label{eq of motion}
\ee
The L.H.S. is a (linearised) Ricci tensor and the R.H.S. can be interpreted 
as the ``matter'' terms.
Then it is easy to see that, in general,  $c_{(n)}(x)\sim 
-M_{(n)}\delta^{(5)}(x-X)$ for some constants $M_{(n)}$,
where $x^i= X^i$ $(i=1,\ldots, 5)$ express the position of 
the rest particle. We thus find that $M_{(n)}$ can read 
as the $n$-loop correction to the  rest mass of our particle.
In this way we can directly evaluate the mass of D-branes from 
the $\beta $-functions.

Especially, it is  easy  to
calculate the 1-loop contribution $c_{(1)}$:
Consider a 1-loop (disk) amplitude  
\be
{\cal A_D} = \int_0^{\infty}\, dT \, \bra{0}e^{-TH}\ket{{\cal D}},
\label{AD}
\ee
where $\ket{{\cal D}}$ denotes the suitable boundary states 
corresponding to the D-brane ${\cal D}$. 
The divergence of moduli integral for 
$ {\cal A_D}$ has  its origin in
the massless components of ${\cal D}$. 
Under the natural assumption for the backgrounds $h_{0\mu}=B_{0\mu}=0$, 
the divergent part of 
$\dsp \frac{\delta}{\delta h_{00}(x)}{\cal A_D}$ is easily calculated as 
\ba
\frac{\delta}{\delta h_{00}(x)}{\cal A_D}&\sim &
 \bra{0}V^{h_{00}}_{(-1,-1)}(x)\, \ket{{\cal D}}^{(0)}_{NS-NS} \nonumber \\
 &=& \dsp - \eta_{00} \frac{\de^{(5)}(x-X)}{\sqrt{\det G_{\mu\nu}}}
 \, \int \, d^p\sigma  \frac{\sqrt{\det (X^*G + X^* B)}}{\sqrt{|G|}}\,\, ,
\label{1-loop}
\ea
where $V^{h_{\mu\nu}}_{(-1,-1)}(x)$ is the graviton emission vertex
in the $(-1,-1)$-picture, and the superscript ``(0)''
indicates the massless sector of the boundary state $\ket{{\cal D}}$.
We should notice that the position  integrals along the Neumann directions
are left (on the other hand, the momentum integrals do not exist for 
these directions).  
Clearly the boundary state of R-R sector does not contribute to 
the above calculation.
Recall the relation $e^{\Phi}= e^{\phi}|G|^{-1/4}$, 
$G_{\mu\nu}=e^{\Phi}g_{\mu\nu}^{(6)}$.
We can also approximate $g_{\mu\nu}^{(6)}$ by the Minkowski metric
$\eta_{\mu\nu}$ in its R.H.S., because  the deviation of metric $h_{\mu\nu}$ 
should be a quantity of the same order as the string coupling.
We can easily obtain 
the 1-loop equation of motion 
\be
   \mbox{\Large$\Box$}h_{00}(x) = e^{-\Phi/2} |G|^{-1/4}
      \int \, d^p\sigma \sqrt{\det (X^*G+ X^* B)}\,\,\delta^{(5)}(x-X)\,\, .   
\label{1-loop 2}
\ee  
Needless to say, this leads to  the same results as those in
the previous sections
\be
\cm_{p}^{(1)}\equiv 
\dil  |G|^{-1/4}\int \, d^p\sigma \sqrt{\det (X^*G+ X^* B)}
\,\,\,.\label{1-loop mass}
\ee
It is no other than the 1-loop mass derived in the previous section.

For higher loop discussions,
it is important
to consider moduli dependent D-brane configurations
and investigate the space-time SUSY breaking from the point of view of
the boundary states. We now focus on only the open string 
loops, since we will later observe that closed string loops are 
negligible in the relevant region for our analysis of D-brane masses. 

It is useful to study  first  the 2-loop case.
Let us take two (the same or different kinds of) D-branes $\ca{D}$, $\ca{D}'$.
The 2-loop amplitude $\cal A_{DD'}$ is nothing but a cylinder.
If there remains any unbroken supersymmetries in the open string channel
\cite{P2}, this amplitude vanishes. 
Therefore the problem if the beta function (\ref{beta function})
has higher loop corrections is reduced to the discussion of the
unbroken space-time supersymmetries.
In the following arguments,
it is convenient to express the boundary
states as
\be
\ket{\ca{D}} = \sum_{s+s'=-2}\ket{\ca{D}}_{s,s'}\,\,,\,\,\,\,
\ket{\ca{D}'} = \sum_{s+s'=-2}\ket{\ca{D}'}_{s,s'}\,\,.
\label{BS}
\ee
The subscripts $s,s'$ express the picture (the ghost charge of 
bosonic ghosts)
and the terms with $s \in \bz$,
$\dsp s \in \frac{1}{2}+\bz$ belong to the NS-NS sector,
the R-R sector respectively.

Let us take a cylinder amplitude with one graviton emission vertex
operator $V^{h_{00}}_{(0,0)}$ in the $(0,0)$-picture,
${\bra{\cal D} V^{h_{00}}_{(0,0)}\ket{{\cal D}'}}$.
It is trivial to extend to the cases with other pictures.
Recall a relation of the graviton vertex operator
$V^{h_{\mu\nu}}_{(0,0)}$
and a photon vertex operator $V^{\mu}_{(0)}$
\begin{eqnarray}
V^{h_{00}}_{(0,0)}&=&V^0_{(0)}\tilde{V}^0_{(0)}\nonumber \\
&=&u_A \tilde{u}_A v_B \tilde{v}_B
\left\{ Q^A_{1/2},\left[ \tilde{Q}^A_{1/2},
V^B_{-1/2}\tilde{V}^B_{-1/2}\right]\right\}\,\,\,,\\
&& u_A \left(-\frac{i}{\sqrt{2}}(C\gamma_\al)^{AB}\right) v_B
  = \de^{0}_{\al} \,\,\,, \label{uv} \\
&&\,\,\,\,(C\,;\,\mbox{charge conjugation matrix with}\,\,\,
C^{-1}\gamma^{\alpha T}C=-\gamma^\alpha )\,\,\,.
\nonumber
\end{eqnarray}
Here the $Q^A_{1/2}$, $\tilde{Q}^A_{1/2}$ are supercharges in
the $1/2$-picture and we used fermion vertex operators
$V^B_{-1/2}$, $\tilde{V}^B_{-1/2}$ in the $-1/2$-picture.
Let us introduce the abbreviated notations such as   
$u\cdot Q \df u_A Q^A_{1/2}, ~ 
v\cdot V \df v_B V^B_{-1/2} $ and so on.
So the above cylinder amplitude can be re-expressed as 
$\dsp 
{\bra{\cal D}
\left\{ u\cdot Q \, ,\left[ \tilde{u}\cdot \tilde{Q},
(v\cdot V) \, (\tilde{v}\cdot \tilde{V}) \right]\right\}
\ket{{\cal D}'}}\,
$
and we can rewrite it 
\be
\begin{array}{l}
\dsp {\bra{\cal D}
\left\{ u\cdot Q \, ,\left[ \tilde{u}\cdot \tilde{Q},
(v\cdot V) \, (\tilde{v}\cdot \tilde{V}) \right]\right\}
\ket{{\cal D}'}}
   \\
\dsp   ~~~~~~~~= - \frac{1}{2}
{\bra{\cal D}
\left\{ u\cdot Q_+(N) \, ,\left[ (N^{-1}\tilde{u})\cdot Q_-(N),
v\cdot V\, (N^{-1}\tilde{v})\cdot  \tilde{V} \right]\right\}
\ket{{\cal D}'}}\,\,, \\
\dsp Q^A_{\pm}(N) := Q^A \pm {N_B}^A \tilde{Q}^B, 
\end{array}
\label{QN}
\ee
for some suitable 
matrix $N$.
For special matrices ${M_B}^A$, ${{M'}_B}^A$ depending
on the moduli fields, the boundary states satisfy relations \cite{CLNY,Li};
\ba
&& Q_+(M)\ket{\cal D}=0 \,\,\,,\nonumber \\
&& Q_+ (M')\ket{{\cal D}'}=0\,\,\,. 
\ea
For a $p$-brane with Neumann coordinates
$\{X^\mu\}$, $(\mu =0,1,2,\cdots ,p)$ with background fields $G$ and
${\cal F}$, the matrix $M$ is written as \cite{CLNY,Li}
\ba
{M_B}^A=i{\left\{\gamma_{01\cdots p}\, \exp\left(\frac{1}{2}
\gamma^{\al \beta}
{\cal F}_{\al\beta}\right) \right\}_B}^A\cdot
\left\{\det (G+{\cal F})\right\}^{-1/2}\,\,\,.\nonumber
\ea
Here the gamma matrices $\gamma_{\al}$, $\gamma^{\al}$ are 
normalized as
\be
\{ \gamma_{\al}, \, \gamma_{\beta} \} = 2 G_{\al\beta}\id , ~~~
\{ \gamma^{\al}, \, \gamma^{\beta} \} = 2 G^{\al\beta}\id ,
\ee 
and 
we introduced the notations for the anti-symmetrized gamma matrices
\be
\gamma_{\al \cdots \beta}:= \gamma_{[\al}\,\cdots \gamma_{\beta ]}\,\,,\,\,\,\,
\gamma^{\al \cdots \beta}:= \gamma^{[\al}\,\cdots \gamma^{\beta ]}\,\,.
\ee
It is well-known that unbroken supersymmetries exist along the directions
$u_A$ such that ${M'}^{-1}M u=u$ (i.e. the eigenvectors of the matrix
${M'}^{-1}M$ with eigenvalue 1): $u\cdot Q_+(M)\ket{\ca{D}}
=u\cdot Q_+(M)\ket{\ca{D}'}=0$,
or equivalently $u\cdot Q_+(M')\ket{\ca{D}}
=u\cdot Q_+(M')\ket{\ca{D}'}=0$ are satisfied.
If there are suitable number of these eigenvectors $u$ so as to 
satisfy the relation \eqn{uv}, 
the cylinder amplitude $\bra{\ca{D}}V^{h_{00}}
\ket{\ca{D}'}$ vanishes because of the identity \eqn{QN}, or equivalently
by the cancellation of NS-NS and R-R sectors. 
It follows  that  the 2-loop corrections to $\beta_{h_{00}}$ should vanish.

On the other hand, in the case that the supersymmetries on the 
D-branes are broken, the cancellation between NS-NS and R-R sectors 
in the amplitudes is not complete and higher loop corrections 
to the $\beta$-function may exist.

Now we study separately 
the aspects of  SUSY breaking in the following two cases:
\begin{enumerate}
\item Only one kind of D-branes $\cal D$ exist and their relative
configurations are all parallel. In this case, all loops of open
string boundaries are the same type of D-branes $\cal D$.
\item Different kinds of D-branes coexist.
\end{enumerate}

~

\noindent
\underline{\em{One kind of parallel D-branes}}

In this case there remains one half of the supersymmetry on the D-brane 
in {\em arbitrary} background moduli and the value of 
the $2$ loop amplitude is zero.
This is because ${M'}^{-1}M \equiv \id$ and so one half the
supersymmetry is trivially preserved.

How about the higher loop cases? 
The higher loop corrections to the $\beta$-function originate from   
the divergence of the amplitudes $\cal A_{D\cdots D}$ when all 
the Teichm\"{u}ller parameters go to infinity. In this situation
the higher loop amplitudes may be factorized to the products of 
the ``pants diagrams'' (or the cylinder amplitudes 
with an insertion of suitable massless vertex). 
Hence  the problem is reduced to the 2-loop case.
We can conclude that all the higher loop corrections 
to the equations of motion vanish thanks to the cancellation by the
supersymmetry.
Therefore the results are exact in one loop level, that is, the analyses
based on the DBI action give rigorous mass spectra!
 (Of course we here neglect the effects of 
closed string loops, and so this observation may be  not correct in the
case when effects of the closed string loops are not small.)

~

\noindent
\underline{\em{Some Different kinds of D-branes}}

Let us take the two different kinds of D-branes  $\ca{D}$, ${\cal D}'$.
(The dimensions of the world volumes of $\ca{D}$ and ${\cal D}'$ 
are different or 
$\ca{D}$ and ${\cal D}'$ are not placed in parallel if they have the same
dimension.)
Here we shall only  consider 
the configurations such that the branes touches  with each other, 
or at least, the branes placed very closely.
This is because we are interested in only bound states of branes
and otherwise, the description by the coordinates of 
the center of mass would lose its meaning.
In this ``intersecting D-brane'' case \cite{D,Se2,BVS2,PT,BL},
we will find out that the cancellation between NS-NS and R-R
sectors is {\em not\,} complete {\em for general background}.

For example, consider a 3-brane wrapping around the 678-th axes 
of $T^4$ and a 1-brane wrapping  around the 9-th axis (Fig.1), which is the 
case we will later analyze in detail. 
It is well-known that this configuration  is supersymmetric
(so-called ``short multiplet'', in which $1/4$ of space-time SUSY are unbroken)
in the special background;
\be
B_{ij}=0 , ~~~~G_{ij}=\left(\begin{array}{cccc}
                            &&&0 \\
                        &      G_{6\leq i,j \leq8}&&0 \\
                            &&&0 \\
                            0&0&0& G_{99}
                          \end{array}   \right).
\label{SUSY config}
\ee
(The 1-brane intersects the 3-brane perpendicularly, and 
the background Kalb-Ramond field is set zero.)
However, if we put general background moduli, 
we have {\em no\/} SUSY any longer.  
In fact, consider the unbroken SUSY charges
$Q_+^A(M)$ and $Q_+^A(M')$ respectively associated with boundary states
$\ket{\cal D}$ and $\ket{{\cal D}'}$. 
In the special background \eqn{SUSY config},
$M'^{-1}M$ indeed has 1 as an eigenvalue. 
But it is not the case for generic moduli. 
It follows that the higher loop corrections 
to $\beta_{h_{00}}$ no longer  vanish. 
This aspect sharply contrasts with the case
of one kind of branes, in which we always have some unbroken SUSY 
{\em independent of the VEV of moduli fields.}       

 \begin{figure}
       \epsfxsize=5cm
       \centerline{\epsfbox{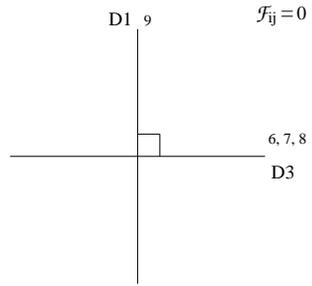}}
       \epsfxsize=5cm
       \centerline{\epsfbox{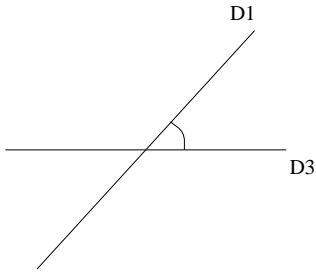}}
       \epsfxsize=5cm
       \centerline{\epsfbox{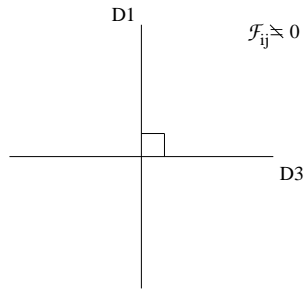}}
       \caption{In the perpendicularly intersecting case with
        ${\cal F}_{ij}=0$, there remains a $1/4$ SUSY. In contrast, the SUSY is
        completely broken in the slantwise intersecting configuration
        or in the  non-vanishing ${\cal F}_{ij}$ case.}
    \end{figure}

We have observed that 
the higher loop corrections to D-brane mass 
is inevitable in the broken SUSY cases.
We will evaluate these  amounts 
in detail in the next subsection.

~

\subsection{Evaluation of Higher-Loop Corrections 
as the Binding Energies}

We explain the method to calculate the contributions from higher
loops to the $\beta$-functions concretely.
The contribution for a fixed diagram comes from the divergent part
of the amplitude when the Teichm{\"u}ller parameters of the world
sheet simultaneously go to large values. Then
the size of all the boundary loops shrink to zero
simultaneously. In this limit only the massless modes are relevant
for our computations.

First of all, we stress the following fact:
We must assume that D-brane mass of 1-loop level, which we evaluated
in the previous section based on the DBI action,  
is sufficiently heavy  for the validity to treat the D-branes 
as static backgrounds. 
In other words, we must consider the cases with large amounts of R-R charges.
Otherwise, we would have to take account of fluctuations 
of the D-brane configurations \cite{KAZA}.
This assumption is also necessary so that the perturbative expansion is 
applicable.  
It is not difficult to show that, under this assumption,
the diagrams with no closed string loops 
are dominant for a fixed number of loops
($L:= \sharp \, \{ \mbox{open string loops}\} \, 
+ 2 \sharp \, \{ \mbox{closed string loops}\} $).
This is because each of the open string boundaries is  assigned a very large 
factor $\sim$ 1-loop mass.
In this way we can conclude 
that only the open string loop diagrams with no genera
are relevant for our calculation.

It is convenient to sum up  all the ``tadpole diagrams'' (Fig.2) first.  
We evaluate the tadpole corrections to the pants-type diagram. 
Each wavy-line connecting a pants and one boundary state expresses the 
graviton and it leads to a factor 
$-\la \hat{\cm}^{(1)}$, which is essentially the normalization 
factor of NS-NS boundary state.  
Here we define the $\hat{\cm}^{(1)}$ as
\be
\hat{\cm}^{(1)}:= e^{\Phi/2} |G|^{1/4} \cm^{(1)} ,
\ee
by using the one-loop mass  $\cm^{(1)}$
evaluated by the DBI action in the previous section.
The reason why it includes $\la$
is clear: Each tadpole gains one  open string  loop.

    \begin{figure}
       \epsfxsize=10cm
       \centerline{\epsfbox{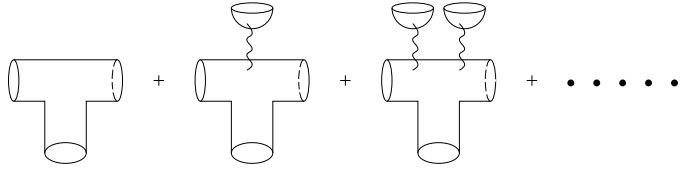}}
       \caption{The bare string coupling is improved by insertions
         of gravitons emitted from boundary states.}
    \end{figure}

We comment on a combinatorial factor for these tadpole diagrams.
When we add one boundary operator to diagrams with $(N-1)$ boundary
operators, there are $N$ cases of possibilities to connect them.
So there are $N!$ cases of possibilities to connect $N$ boundary
operators and a bare coupling pants diagram each other.
But there is a symmetry factor $1/N!$ for these diagrams because of 
the invariance under the permutation 
of $N$ boundaries, and this cancels the factor $N!$.

In this way the summation of all the tadpole diagrams 
leads to the following factor;
\be
1+ (-\lambda \hat{\cm}^{(1)}) +(-\lambda \hat{\cm}^{(1)})^2+
(-\lambda \hat{\cm}^{(1)})^3+\cdots = \frac{1}{1+\lambda \hat{\cm}^{(1)}}
\sim \frac{1}{\lambda \hat{\cm}^{(1)}}\,\,\,.
\ee
In the last line, we used our assumption that the one-loop D-brane mass
is very large.
We may also reinterpret this correction as follows: The string 
coupling $\lambda$ should be  replaced with an ``effective string coupling''
$\dsp \la_{\msc{eff}} \equiv \frac{\la}{\la \hat{\cm}^{(1)}}
\equiv \frac{1}{\hat{\cm}^{(1)}}$,
by taking account of all the tadpole diagrams.
However  one should notice a fact: Generally,
in the limit of large
 Teichm\"{u}ller parameters,  any $L$-loop diagram can be factorized into 
($L-1$) pants diagrams. On the other hand, we know an $L$-loop diagram 
should include a factor of $\la^{L-2}$. So, we find that 
any $L$-loop diagram should include a factor 
\be
\la^{L-2}\,\left( \frac{1}{\la \hat{\cm}^{(1)}}\right)^{L-1} =
\la^{-1}\, \left( \frac{1}{\hat{\cm}^{(1)}}\right)^{L-1}\,\,\,.
\label{tadpole factor}
\ee

We make one remark here:
When we consider an arbitrary diagram with odd-number of open boundary
loops, the amplitude necessarily contains tadpole type amplitudes.
But we have already included  all the contributions from tadpole type
diagrams (Fig.3) into 
the factor \eqn{tadpole factor} and it will be an over
counting if we take these diagrams into account.

Collecting all the above observations, we can conclude that only the
diagrams with the following two properties 
can contribute to the calculations:

\begin{enumerate}
\item A diagram with no closed string loops.
\item A diagram with no tadpole, which is inevitably 
      a diagram with only even-number of open string loops.
     The contributions of the tadpole type diagrams are already 
    included in the factor \eqn{tadpole factor}.
\end{enumerate}

    \begin{figure}
       \epsfxsize=10cm
       \centerline{\epsfbox{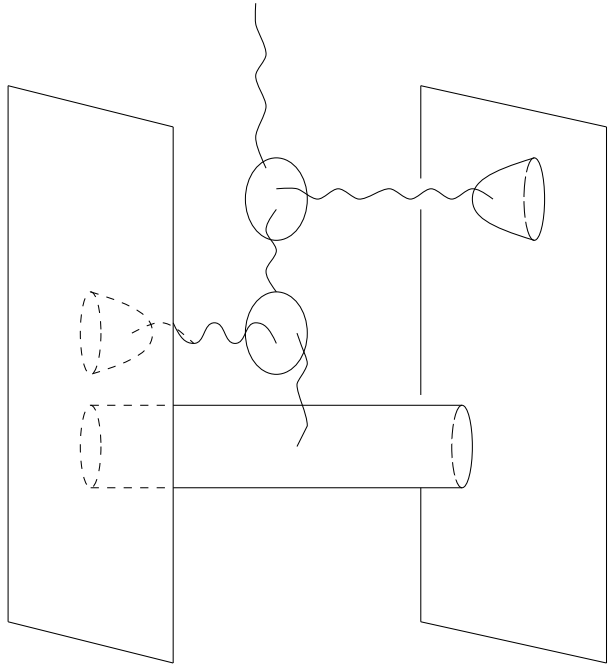}}
       \caption{Contributions from diagrams with tadpoles are already
          included in the effective string coupling.}
    \end{figure}

    \begin{figure}
       \epsfxsize=10cm
       \centerline{\epsfbox{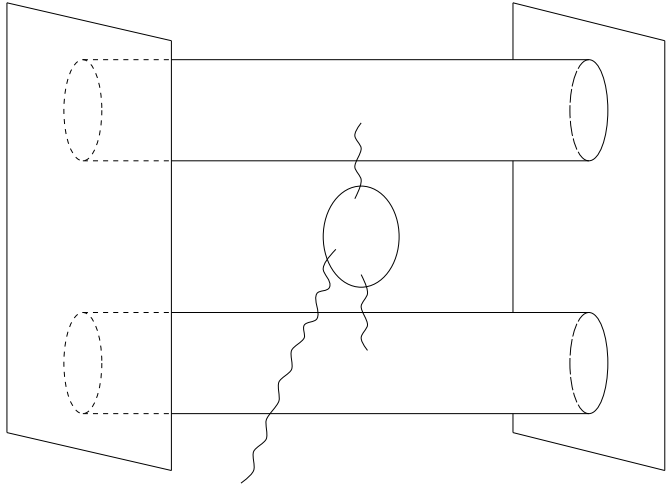}}
       \caption{Only this type of diagrams contribute to the beta
         function calculation.}
    \end{figure}

As a result, there are contributions to the $\beta$-function from 
diagrams composed of only cylinder type diagrams whose two boundaries
are put on different D-branes (Fig.4).

~

Now, we arrive at the stage to evaluate concretely the higher loop 
corrections to the D-brane masses.
As a simple example,
let us consider the type IIB case compactified on $T^4$ with
coordinates $(x^6,x^7,x^8,x^9)$ and take  an
``Intersecting D-brane'' configuration of Dirichlet 1-branes and
D3-branes.
Consider $n$ D1-branes (${\cal D}$) wrapping around the $9$th 
axis and $n'$ D3-branes (${\cal D}'$) wrapping around the $678$th axes. 
The center of mass of this system is specified by 
$(X^1,X^2,\cdots ,X^5)$.

The one-loop mass $\cm_{1-3}^{(1)}$ was calculated from the disk amplitude 
$c^{(1)}$ in section 3
\ba
\cm_{1-3}^{(1)}&\equiv & 
e^{-\Phi/2}|G|^{-1/4}\hat{\cm}_{1-3}^{(1)}\,\,\,, \label{M1-1} \\
\hat{\cm}_{1-3}^{(1)}&=& \dsp n \sqrt{G_{99}}+n'
\sqrt{|G| \left\{G^{-1}-(\ast B)G (\ast B)\right\}^{99}}\,\,\,, \label{M1} \\
{(\ast B)}^{ij} &:=&{\dis \frac{1}{\sqrt{|G|}}}\cdot
\frac{1}{2}\epsilon^{ijkl}B_{kl}\,\,\, . \nonumber
\ea

First we take a $2$-loop cylinder diagram whose boundaries are put
on the D1-brane ${\cal D}$, the D3-brane ${\cal D}'$, respectively. 
As we already pointed out, for the computation of $\beta$-functions
we only have to evaluate the value of amplitude in the limit of 
large Teichm\"{u}ller parameters, 
which is the IR limit in the closed string channel 
(or equivalently, UV limit in the open string 
channel).
In this limit only the massless components of boundary states can 
contribute
\be
c^{(2)} \sim \, {}^{(0)}\bra{\ca{D}}V^{h_{00}}\ket{\ca{D}'}^{(0)},
\ee
where the superscript $(0)$ indicates the massless part.
However, there is one subtle point. The massless condition 
indicates that the momenta of particles which are exchanged 
between the boundaries must be zero. (Under the Neumann boundary 
condition, one can generally obtain non-zero winding. 
In our case, because the 0-th direction
is Neumann and non-compact,
the winding should be also 0. So, we obtain $p^0_L=p^0_R=0$.)
If considering on the conservation of the ghost charge, 
we can find that a closed string propagator should be inserted.
It is the result of an integral about the position of an 
inserted vertex $V^{h_{00}}$. 
Naively the insertion of  this propagator seems to lead 
to a divergence because of the zero-momentum. But an appropriate
choice of picture of $V^{h_{00}}$ gives a momentum factor which cancels
the pole of the propagator. 
More precisely, we should put some infinitesimal 
momenta along the 1$\sim$5th directions  
to the vertex operator $V^{h_{00}}$.
After the calculation they should be set
zero. (Notice that for the compact 6,7,8,9th directions the momenta 
are quantized and the 0th direction is always Neumann.)
In the calculation of this type, we obtain contributions 
only from the ``contact terms'' among  $V^{h_{00}}$ and 
the shrinked open boundaries on D-branes.

It is convenient to  make use of the same technique as \eqn{QN};
\ba
{}^{(0)}\bra{\ca{D}}V^{h_{00}}\ket{\ca{D}'}^{(0)} &=& 
{}^{(0)}{\bra{\cal D}
\left\{ u\cdot Q \, ,\left[ \tilde{u}\cdot \tilde{Q},
v\cdot V\, \tilde{v}\cdot \tilde{V} \right]\right\}
\ket{{\cal D}'}}^{(0)}   \label{2loop factor} \\
   &=&\dsp - \frac{1}{2}
{}^{(0)}{\bra{\cal D}
\left\{ u\cdot Q_+(M) \, ,\left[ (M^{-1}\tilde{u})\cdot Q_-(M),
v\cdot V\, (M^{-1}v) \cdot \tilde{V} \right]\right\}
\ket{{\cal D}'}}^{(0)} \label{QM}\,\,,  \nonumber \\
  M &:=& i \frac{\gamma_0 \gamma_9}{\sqrt{-G_{00}G_{99}}}, 
\nonumber
\ea
where $M$ is chosen so that $Q_+(M)\ket{\ca{D}}^{(0)}=0$.
Since ${}^{(0)}\bra{\ca{D}}Q_+(M) =0$, 
the most non-trivial part of the calculation is to evaluate
the term $Q_+(M)\ket{\ca{D}'}^{(0)}$.
Making use of the identity
\be
\begin{array}{l}
Q_+(M')\ket{\ca{D}'}^{(0)}= 0 , ~~~
\dsp M' := i \frac{e^{\frac{1}{2}\tilde{\ca{F}}_{\al\beta}\gamma^{\al\beta}}}
{\sqrt{-G_{00}}\sqrt{\det (\tilde{G}+\tilde{\ca{F}})}}\gamma_0 
\gamma_{678}\,\,,\\
\dsp \tilde{G}\equiv G_{6 \leq i,j \leq 8}\,\,,~~~~
\tilde{\ca{F}}\equiv \ca{F}_{6 \leq i,j \leq 8}\,\,,
\end{array}
\label{QM'}
\ee
and taking the spinor $u$ with a property; 
\be
\frac{1}{\sqrt{|G|}}\gamma_{6789}u = -u ,
\label{susy u}
\ee  
we obtain
\ba
 u\cdot Q_+(M)\ket{\ca{D}'}^{(0)}& =& \frac{\sqrt{|G|}\tri}{\ca{N} \ca{N}'} 
u \cdot Q \ket{\ca{D}'}^{(0)} + \cdots , \label{2loop factor2} \\
\tri & :=&
\sqrt{G_{99}}\sqrt{\left\{G^{-1}-(\ast B)G(\ast B)\right\}^{99}}
-\left(1-G_{9i}\cdot (\ast B)^{i9}\right)
\label{triangle}\,\,,\\  
\ca{N}&:=& \sqrt{G_{99}}\,\, , \nonumber \\
   \ca{N}' &:=& \sqrt{\det (\tilde{G}+ \tilde{\ca{F}})}\nonumber  \\
           & \equiv& 
   \sqrt{|G| \left\{G^{-1}-(\ast B)G(\ast B)\right\}^{99}}\,\,. \nonumber
\ea
In the R.H.S. of above identity \eqn{2loop factor2}, 
``$\cdots $'' denotes the unimportant terms that do not contribute to our 
calculation.
$\ca{N}$, $\ca{N}'$ are no other than the normalizations of 
the boundary states in NS-NS sector $\ket{\ca{D}}$, $\ket{\ca{D}'}$.
The spinor $u$ satisfying \eqn{susy u} actually exists, since 
$\dsp \left(\frac{1}{\sqrt{|G|}}\gamma_{6789}\right)^2 = \mbox{\bf{1}}$ 
holds, and we should remark the fact that  $u$ does not depend on any moduli.
It is further worth while to note that $ u\cdot Q_+(M)$ under 
\eqn{susy u} is no other than  the unbroken SUSY charge 
in the supersymmetric configuration of branes \eqn{SUSY config}. 
Under \eqn{SUSY config},
we obtain $ u\cdot Q_+(M) \ket{\ca{D}'} =0 $, and especially 
$\tri = 0$. This choice of $u$ is appropriate to our calculations, 
because we are now interested in the perturbative 
expansions around the supersymmetric vacuum\footnote
   {If we consider the bound states of D-branes and {\em anti\/}-D-branes,
    namely, in the case $u\cdot Q_+(M) \ket{\ca{D}} =0, ~
      u\cdot Q_-(M') \ket{\ca{D}'} =0
    $, we should take $u$ such that $\frac{1}{\sqrt{|G|}}\gamma_{6789}u = u$
     for correct perturbative calculations.}. 
It may be natural to regard $\tri$
as a value characterizing SUSY violation.

Now it is not hard to write down the 2-loop correction to $\beta_{h_{00}}$.
After including all the tadpole corrections 
(\ref{tadpole factor})(\ref{2loop factor})(\ref{2loop factor2}), we can
obtain 
\ba
c^{(2)}(x) &=&  \frac{1}{2}\, G_{00} \la^{-1}\, \frac{1}
                                  {\hat{\cm}_{1-3}^{(1)}} 
      (2 nn'\sqrt{|G|} \tri) \, 
      \frac{\delta^{(5)}(x-X)}{\sqrt{\det G_{\mu\nu}}\, \sqrt{|G|}} 
 \nonumber \\
&=&   \eta_{00} \, \frac{nn' \tri}{ (|G|^{-1/4}\hat{\cm}_{1-3}^{(1)})} \, 
    \frac{\delta^{(5)}(x-X)}{\sqrt{\det G_{\mu\nu}} \, \sqrt{|G|}}\,\, .
\label{2-loop}
\ea
In the first line of (\ref{2-loop}),
the factor 2  before $nn' \tri$ corresponds to 
the existence of contact terms between the graviton vertex and 2 boundaries,
one of which  resides on the 1-branes ${\cal D}$ 
and the other of which does on the 3-branes ${\cal D}'$. 
The factor $\frac{1}{2}$ is nothing but a symmetry factor
due to the exchange of the boundaries of cylinder.  
Thus the 2-loop correction to the mass can be read as follows;
\be
\cm_{1-3}^{(2)} = - e^{-\Phi/2} \, 
\frac{nn' \tri}{ (|G|^{-1/4}\hat{\cm}_{1-3}^{(1)})}\,\,. 
\label{2-loop mass}
\ee 
It implies that we may naturally regard  
$\tri$ as  the binding energy
between two types of D-branes ${\cal D}$, ${\cal D}'$.
As is already observed, $\tri$ vanishes
under the supersymmetric brane configuration \eqn{SUSY config},
but does not for the general non-supersymmetric backgrounds. 
We can conclude that {\em the binding energy among the branes 
has its origin in the SUSY violation.}

    \begin{figure}
       \epsfxsize=10cm
        \centerline{\epsfbox{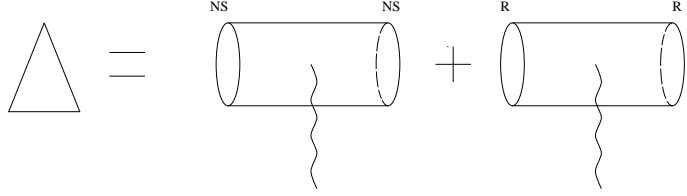}}
         \caption{The cylinder amplitude $\triangle$
          appears in the 2-loop beta function calculation. 
          In the ``Intersecting D-brane'' case, the cancellation 
          between the NS-NS and the R-R does not exist.}
    \end{figure}

Next we consider the contributions from $2k$ $(k\geq 2)$ loop 
$c^{(2k)}$. As is already commented, in the limit that  all the areas of 
 open string boundaries shrink, only  the diagrams of the type in figure 4
survive, and they are factorized into the $k$ products 
of cylinder amplitudes of the forms
${}^{(0)}\bra{\ca{D}}V^{h_{\mu\nu}}\ket{\ca{D}'}^{(0)}$.
Hence we obtain
\ba
c^{(2k)}(x) &=& \mbox{combinatorial factor}\times 
G_{00} \la^{-1}\,\frac{ (2 nn'\sqrt{|G|} \tri)^k}
                  {(\hat{\cm}_{1-3}^{(1)})^{2k-1}} 
      \, \frac{\delta^{(5)}(x-X)}{\sqrt{\det G_{\mu\nu}}\, \sqrt{|G|}} 
\nonumber \\
&=& \mbox{combinatorial factor}\times 
\eta_{00}\,\frac{ (2 nn'\tri)^k}{(|G|^{-1/4}\hat{\cm}_{1-3}^{(1)})^{2k-1}} 
      \, \frac{\delta^{(5)}(x-X)}{\sqrt{\det G_{\mu\nu}}\, \sqrt{|G|}}  
\,\,\,.\label{2k loop}
\ea

Let us evaluate  the combinatorial factor of the diagram.
When we add one cylinder
$\sim {}^{(0)}\bra{\ca{D}}V^{h_{\mu\nu}}\ket{\ca{D}'}^{(0)}$
to diagrams with $2(k-1)$ open string loops,
there are ($2k-3$) cases of possibilities to connect them.
So there are $(2k-3)!!$ diagrams with a fixed loop number $2k$.
(Note that adding one cylinder
$\sim {}^{(0)}\bra{\ca{D}}V^{h_{\mu\nu}}\ket{\ca{D}'}^{(0)}$ 
needs another pants diagram connecting to the original diagram.)
But one more symmetry factor $\dsp \frac{1}{2^k\, k!}$ 
for these diagrams appears.
These lead to a correct (moduli-independent) numerical coefficient 
${\dis \frac{(2k-3)!!}{2^k \, k!}}$.
So, the $2k$-loop correction to the mass $\cm_{1-3}^{(2k)}$ is given by 
\be
\cm_{1-3}^{(2k)} = - \frac{(2k-3)!!}{k!}
e^{-\Phi/2} \, \frac{\left(nn' \tri\right)^k}
{ (|G|^{-1/4}\hat{\cm}_{1-3}^{(1)})^{2k-1}}\,\,\,. 
\label{2k-loop mass}
\ee 
 
Collecting the results 
(\ref{M1-1})(\ref{2-loop mass})(\ref{2k-loop mass}), 
we can finally get  the mass formula  for
this bound state of the  intersecting $n$ D1-branes and $n'$ D3-branes
including all the corrections of higher loops;
\ba
\cm_{1-3}&=& \cm_{1-3}^{(1)} \left[1- (n n' \tri)
\left(\frac{1}{|G|^{-1/4}\hat{\cm}_{1-3}^{(1)}}\right)^2
-{\dis \sum^{\infty}_{k=2}\frac{(2k-3)!!}{k!}}
\,(n n' \tri)^k\,
\left(\frac{1}{|G|^{-1/4}\hat{\cm}_{1-3}^{(1)}}\right)^{2k}
\right]\nonumber \\
&=& \cm_{1-3}^{(1)}
\sqrt{1-\frac{2 n n' \tri}{(|G|^{-1/4} \hat{\cm}_{1-3}^{(1)})^2}}\,\,\,.
\label{mass formula}
\ea
This result  exactly reproduces the BPS mass formula
predicted by U-duality in section 2!

~

Until now we only consider  a bound state of the intersecting 
D1- and D3-branes in the type IIB string  compactified over $T^4$. 
The applications to other bound states are straightforward.
For a bound state of $n $ Dirichlet $0$-branes and $n'$ D4-branes
in the $T^4$-compactified type IIA theory, we only have to replace the
value of $\tri$ with
\ba
\tri = 
\sqrt{1 + \frac{1}{|G|^{1/2}}\left( \int_{T^4}B\wedge\ast B\right)
+ \frac{1}{4|G|} 
\left( \int_{T^4}B\wedge B\right)^2}
- \left( 1- \frac{1}{2|G|^{1/2}} \int_{T^4} B\wedge B \right)\,\, .
\nonumber
\ea 
This result is also consistent with the prediction of M-theory
in Eq.(\ref{D0D4mass}).

Next let ${\cal D}$, ${\cal D}'$ be respectively, $n$ Dirichlet $2$-brane
wrapping around $67$th directions and $n'$ D2-brane wrapping around $89$th 
axes in the type IIA theory compactified on the $T^4$.
For a bound state of ${\cal D}$ and ${\cal D}'$,
we obtain its binding energy 
\ba
\tri &=& |G|^{-1/2}
\sqrt{\sum^{3}_{i=1} \left(\int_{\Sigma_{1}} X^{\ast} J_i \right)^2 +
\left(\int_{\Sigma_{1}} X^{\ast} B \right)^2}
\, 
\sqrt{\sum^{3}_{i=1} \left(\int_{\Sigma_{2}} X^{\ast} 
J_i \right)^2 +
\left(\int_{\Sigma_{2}} X^{\ast} B \right)^2} \nonumber  \\
 & & \hspace{3cm}-\left\{ 1-(G_{68}G_{79}-G_{69}G_{78})- B_{67}B_{89}\right\},
\label{tri 2-2}
\ea 
where the $\Sigma_1$ and $\Sigma_2$ are respectively the world volumes of the
D2-branes ${\cal D}$ and ${\cal D}'$.

We make one remark for this intersecting D2-brane case.
Let us assume $n=n'$.
As we showed in section 3, 
by considering general 2-cycle $S \in H_2(T^4;\bz)$ 
(perhaps, having a higher genus), 
the analysis based on the DBI action gives
a correct mass formula for 2-branes {\em including general bound states}.
Why did  it give a correct answer in spite of the restricted analysis
only in the one-loop  we have done there?
In the study in this section, we took the constant background field
and discussed higher loop corrections.
In contrast, in section 3 we considered 
holomorphic embeddings with the maps $\{X\}$'s
directly. Then the induced background fields 
$X^{\ast}G_{ij}$, $X^{\ast}B_{ij}$
on the world volume are {\em not\/} constant in general.
These non-constant holomorphic embeddings are known as 
supersymmetric cycles
and guarantee the cancellation of higher (more than one) loop
corrections to the $\beta$-function.
That is the reason why we obtained the correct mass formula
only in the one-loop analysis for the D2-brane case.

In summary, in order to treat the bound states of the two kinds of 
D2 branes, one half of which is wrapping around $T^{67}$ and the
other half of which is wrapping around  $T^{89}$,
we can incorporate all the quantum corrections
(perturbative loop corrections) into a single Dirac-Born-Infeld action
with a genus ``two'' world volume. That is a homological sum of 
$n$ holomorphic curves 
(supersymmetric cycles) with genus two (Fig.6).
    \begin{figure}
       \epsfxsize=10cm
       \centerline{\epsfbox{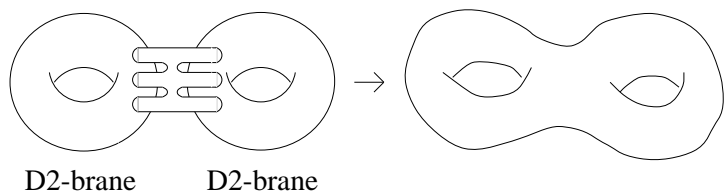}}
       \caption{The bound state of intersecting D2-branes can be
       represented as a SUSY cycle (holomorphic curve) with genus two.}
    \end{figure}
It may be plausible to interpret this aspect as a ``{\em geometrization of 
quantum corrections\/}''. 

On the other hand,
for a bound state of the D1-branes and D3-branes, the dimensions of these
two kinds of branes are different and we cannot describe the state by a
single DBI action (Fig.7).
    \begin{figure}
       \epsfxsize=10cm
       \centerline{\epsfbox{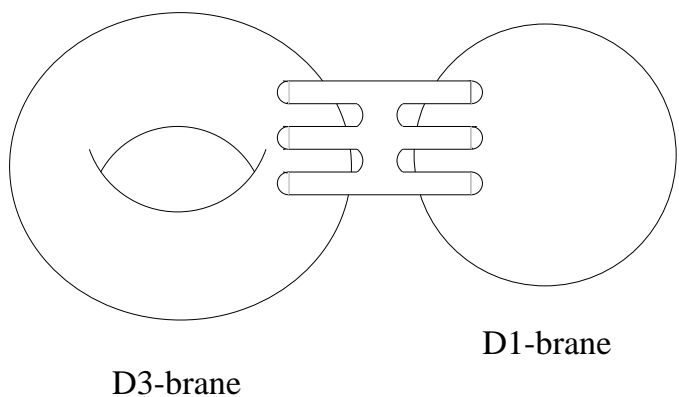}}
       \caption{The dimensions of the D3-brane and the D1-brane are
     different and we cannot represent 
     its bound state as a single world volume.}
    \end{figure}

Another interesting example of the ``geometrization'' is the situation 
of the {\em branes within branes} \cite{D}. Let us again consider
the bound states of D1-, D3-branes in IIB string. 
But, this time we make  $n$ D1-branes wrap around the 6th-axis and  
$n$ D3-branes wrap around the 678th-axes. We further assume that 
each 1-brane is contained in each 3-brane.
Clearly this configuration of branes breaks SUSY completely 
(in {\em any\/} background!), and we have a non-zero binding energy
\be
\tri = \sqrt{G_{66}}\sqrt{\left\{G^{-1}-(\ast B)G(\ast B)\right\}^{99}}
+ G_{6i}\cdot (\ast B)^{i6}. 
\ee
However,
we already knew this case can be also described by a single DBI action
with a suitable choice of a background gauge field:
$$
\ca{F}\equiv F+B, ~~~~ [F] =\mbox{
Poincar\'{e} dual of the 1-brane in the world volume of the 3-brane.}
$$
This  is a  manifestly supersymmetric treatment.
It is straightforward to check that these two treatments yield
the same mass formula.
Therefore, we can find out a remarkable fact: In the case of {\em branes 
within branes},
the non-supersymmetric calculation with non-zero higher loop
corrections is equivalent to the supersymmetric treatment based on 
the single DBI action with suitable background gauge fields.
That is, all loop corrections can be transmuted into  
a charge of the gauge field!
One may say this is another example of the geometrizations of quantum
corrections.

In the relation to this subject, it may be also meaningful to discuss 
the non-abelian extension when the gauge symmetry enhancement occurs on 
D-branes. In section 3, we only considered the charges of background 
gauge fields along the ``$U(1)$-sectors''. This approach was limited in
the sense that we can only realize as the  ``$U(1)$-charges''
the brane configurations that can be 
reduced to one kind of  branes by T-duality. 
However, at least in the level of naive observation, 
if the gauge theory on branes becomes non-abelian, 
we can interpret more general configurations, 
which is not necessarily reduced to one kind of 
branes by T-dualiy, as the characteristic classes composed of the field 
strength. The non-abelian extension of DBI action is proposed by 
Tseytlin \cite{NBI}, in which the symmetrized traces of the products 
of field strength appear. 
It may be interesting to check the consistency between the two general 
descriptions of bound states, one of which is based on the non-abelian
Born-Infeld action and the other of which is based on the string loop
analysis given in the present  section. 

To close this section we again emphasize that 
the DBI action (even the non-abelian DBI)
is not sufficient to describe {\em all\/} the bound states.
We must inevitably perform the higher loop analysis to complete
our studies.

~

~


\section{Conclusions and Discussions}
\cleqn

In this paper, we investigated the mass spectra of R-R solitons
by making use of the D-brane techniques in order to confirm the U-duality.
We would like to emphasize that our results are obtained under the completely
general backgrounds. Especially it is remarkable that the form of the DBI
action is perfectly fitted to the moduli dependence of the masses of 
BPS solitons with (one kind of) the R-R charges.  
Moreover, the masses of some bound states 
- {\em branes within branes\/} -
can be also evaluated by the DBI action by
incorporating suitable charges of gauge fields.
In other words, we have shown that  
the DBI action is consistent with one of the T-duality transformations - 
``integer theta parameter shift'' $B_{ij}\rightarrow B_{ij} +\Theta_{ij}$.

It is a challenging task to analyse more general bound states.
We argued on these states, emphasizing the relation with the SUSY violation,
and observed that the 
quantity $\tri$ which characterizes SUSY violation
can be interpreted as  a binding energy of D-branes.   
The characteristic $\tri$ is essentially a sum of contributions from
annulus amplitudes in both NS-NS sector and R-R sector.
When cancellations between two sectors in these amplitude are not
complete, the 
space-time SUSY is broken and
 there exist binding energies among branes.
In the case that the only open string loop configurations are dominant, 
we evaluated the contributions to the binding energy from all these loops.
We have shown that the summation of all loop corrections yields the consistent
results with U-duality.

However, there are still some open problems.
In order to complete our D-brane analysis of BPS mass spectra,
we have to further consider three open problems which we presented  
in the opening of section 4; The first is the
analysis of the bound states of D-branes and 
the fundamental excitations. The second is the
 evaluation of the dependence of BPS masses on 
the R-R moduli. The problem 
about the extra R-R charge \cite{GHM} originated from 
the 1st Pontrjagin number of K3 is the third one.
We believe that these problems can be also solved by higher loop analyses 
to $\beta$-functions. 
In section 2 we discussed the BPS mass formulae based on some part of  
U-duality invariance. However,  according to the analysis in M-theory, 
we can write down more complete mass formula Eq.(\ref{susyMass}) 
(with the suitable replacements for the charges $q$, $r_l$, $s^{kl}$
by $\tilde{q}$, $\tilde{r}_l$, $\tilde{s}^{kl}$). 
This has the full U-invariance $O(5,5;\bz)$ and remarkably, it   
includes some interaction terms between the fundamental excitations
and R-R solitons. 
Therefore, the first problem is
 especially significant in order to confirm the full U-duality 
from the viewpoints of the D-brane analysis and also to check the consistency
between the D-brane calculations and the M-theory approach.
We wish to present a more detailed study on this subject in future.
In the last problem, 
we will have to treat carefully 
the open string loops in the twisted sectors of the K3 orbifold.

It is also worth while remarking  on the closed string loops, which 
we neglected in the discussion of section 4. 
Our analysis with only open string loops is valid in
the cases when the D-branes are very heavy and there are no recoils
between them. In other words, these are
the cases that R-R charges $N$ assigned to the D-branes are  
very large and we treat them (semi)classically 
(i.e. we do not consider the quantum fluctuations of the
branes). This large $N$ case is the situation that M(atrix) theory
 \cite{BFSS,IKKT,DVV} has its mean 
as a M-theory in the infinite momentum (light-cone) frame.
This M(atrix) theory is realized as a large $N$ limit of a SUSY
Yang-Mills theory.

Now, there is a curious point to be mentioned 
in the relation with the recent studies about M(atrix) theory.
In the works \cite{B2,PP} the calculations 
of quantum corrections (loop corrections and instanton corrections)
in the SUSY Yang-Mills 
on the world brane are compared with the {\em tree\/}  
calculation in (11D) bulk SUGRA and they claim these are equivalent.
However, there is a naive question: 
How about the quantum corrections in SUGRA? 
If we assume the description by the 
M(atrix) theory is completely valid, the consistency 
of the computations in \cite{B2,PP}
will demand that,  in the large $N$-limit, the tree level of bulk SUGRA 
should be exact.
As a result, this {\em classical\/} SUGRA will become 
equivalent to the {\em quantum\/} SYM in this limit.

On the other hand, in this paper
we evaluated  the BPS mass formulae from the open stringy 
loop corrections under  the D-brane backgrounds 
and compared the results with the mass formulae
obtained by the  classical  SUGRA (U-duality).
We have actually observed  that 
the closed string loop corrections can be neglected  
in the limit of large R-R charges.
Recalling  the fact that the loop corrections in open string theory 
correspond to those in SYM and the closed string loops correspond to 
those in SUGRA in the low energy limit, 
our results seem to support the validity of M(atrix) theory! 
Our analysis is still limited, but
we hope it will give some insights  to the studies of M(atrix) theory
in future.

Although the above consideration is satisfactory,
it may be still meaningful to ask 
whether the closed string loop corrections {\em exactly\/}
vanish, because the U-duality should be valid even if the amount of 
R-R charges $N$ is a small value. 
One possibility that the contributions from closed string loops
do not break our analysis even in the cases with small R-R charges
is a Fischler-Susskind type mechanism \cite{FS}. 
Namely, all the closed string loop corrections might contribute to only  
the renormalization of dilaton ($=$ string coupling constant),
and hence the mass formulae might be kept essentially unchanged.
However, it remains an open problem 
for a long whether this mechanism can 
apply to supersymmetric theories in higher loop
order when supersymmetry is broken by boundary conditions.
Anyway, we will have to treat carefully the quantum fluctuations
of D-branes as in the discussions in \cite{KAZA} 
in order to work  properly in the region where R-R charges
are not large. 

~

~

\section*{Acknowledgement}
We are especially grateful to  H. Ishikawa and Y. Matsuo who participated
in the early stage of this work. Y.S. also thanks to T. Eguchi for helpful
discussions. K.S. thanks to K.~Ezawa for useful comments.

\end{document}